\documentclass[journal]{IEEEtran}
\usepackage{cite}
\usepackage{mathtools}
\usepackage{amsmath,amssymb,amsfonts}
\usepackage{graphicx}
\usepackage{textcomp}
\usepackage{stfloats}
\usepackage{xcolor}
\usepackage{gensymb}
\usepackage{algorithm}
\usepackage{algorithmic}
\usepackage[acronyms,nonumberlist]{glossaries}
\usepackage{multirow}
\usepackage{array}

\usepackage{adjustbox}
\makeatletter
\newcommand\Label[1]{&\refstepcounter{equation}(\theequation)\ltx@label{#1}&}
\makeatother

\def\BibTeX{{\rm B\kern-.05em{\sc i\kern-.025em b}\kern-.08em
    T\kern-.1667em\lower.7ex\hbox{E}\kern-.125emX}}

\newcolumntype{C}[1]{>{\centering\let\newline\\\arraybackslash\hspace{0pt}}m{#1}}

\newtheorem{remark}{Remark}

\newacronym{BS}{BS}{base station}
\newacronym[plural=UEs,
            longplural={user equipment}]{UE}{UE}{user equipment}
\newacronym{UL}{UL}{uplink}
\newacronym{DL}{DL}{downlink}
\newacronym{LoS}{LoS}{line-of-sight}
\newacronym{NLoS}{NLoS}{non-line-of-sight}
\newacronym{RIS}{RIS}{reconfigurable intelligent surface}
\newacronym{SD}{SD}{subsurface design}
\newacronym{ESD}{ESD}{extended subsurface design}
\newacronym{ISD}{ISD}{iterative subsurface design}
\newacronym{CISD}{CISD}{converged ISD}
\newacronym{MU}{MU}{multi-user}
\newacronym{SU}{SU}{single-user}
\newacronym{MF}{MF}{matched filtering}
\newacronym{SNR}{SNR}{signal-to-noise ratio}
\newacronym{SINR}{SINR}{signal-to-interference-plus-noise ratio}
\newacronym{CSI}{CSI}{channel state information}
\newacronym{SVD}{SVD}{singular value decomposition}
\newacronym{PDF}{PDF}{probability density function}
\newacronym{AWGN}{AWGN}{additive white Gaussian noise}
\newacronym{SC}{SC}{scattered channel}
\newacronym{VURA}{VURA}{vertical uniform linear array}
\newacronym[plural=AoAs,
            longplural={angles-of-arrival}]{AoA}{AoA}{angle-of-arrival}
\newacronym[plural=AoDs,
            longplural={angles-of-departure}]{AoD}{AoD}{angle-of-departure}
\newacronym{TMSE}{TMSE}{total mean-squared error}
\newacronym{MMSE}{MMSE}{minimum mean-squared error}
\newacronym{MSE}{MSE}{mean-squared error}
\newacronym{BW}{BW}{bandwidth}
\newacronym{RV}{RV}{random variable}
\newacronym{AI}{AI}{artificial intelligence}
\newacronym{ML}{ML}{machine learning}
\newacronym{BER}{BER}{bit error rate}
\newacronym{OFDM}{OFDM}{orthogonal frequency-division multiplexing}
\newacronym{MIMO}{MIMO}{multiple-input-multiple-output}
\newacronym{IUI}{IUI}{inter-user interference}

\begin{document}

\title{Phase Selection and Analysis for Multi-frequency Multi-user RIS Systems Employing Subsurfaces in Correlated Ricean and Rayleigh Environments}
\author{Amy~S.~Inwood,~\IEEEmembership{Member,~IEEE,}       
        Peter~J.~Smith,~\IEEEmembership{Fellow,~IEEE,}
        Philippa~A.~Martin,~\IEEEmembership{Senior Member,~IEEE,}
        and~Graeme~K.~Woodward,~\IEEEmembership{Senior Member,~IEEE}

\thanks{A. S. Inwood was with the Department of Electrical and Computer Engineering, University of Canterbury,
    Christchurch, New Zealand, and is now with the Centre for Wireless Innovation (CWI), Queen’s University Belfast, Belfast, BT3 9DT, U.K. (email: a.inwood@qub.ac.uk).}
    \thanks{P. J. Smith is with the School of Mathematics and Statistics, Victoria University of Wellington, Wellington, New Zealand (e-mail: peter.smith@vuw.ac.nz).}
    \thanks{P. A. Martin is with the Department
    of Electrical and Computer Engineering, and G. K. Woodward is with the Wireless Research Centre, University of Canterbury,
    Christchurch, New Zealand (e-mail: (philippa.martin, graeme.woodward)@canterbury.ac.nz).}}
\maketitle

\begin{abstract}
Phase selection design for reconfigurable intelligent surfaces (RISs) is a significant research challenge, as a closed-form optimal solution for a multi-user (MU) system is believed to be intractable. While existing methods achieve strong near-optimal performance, they typically entail high computational complexity. In this work, we take a different approach and propose a practical method that achieves competitive performance while substantially reducing computational complexity. To do so, we consider a RIS divided into subsurfaces. Each subsurface is designed specifically for one user, who is served on their own frequency band. The other subsurfaces (those not designed for this user) provide additional uncontrolled scattering. We derive the exact closed-form expression for the mean signal-to-noise ratio (SNR) for the proposed subsurface design (SD) when all channels experience correlated Ricean fading. We simplify this to find the mean SNR for line-of-sight (LoS) channels and channels experiencing correlated Rayleigh fading. An iterative SD (ISD) process is proposed, where subsurfaces are designed sequentially, and the phases that are already set are used to enhance the design of the remaining subsurfaces. This is extended to a converged ISD (CISD), where the ISD process is repeated multiple times until the SNR increases by less than a specified tolerance. The ISD and CISD both provide a performance improvement over SD, which increases as the number of RIS elements increases. The SD is significantly simpler than the lowest complexity MU method we know of, and despite each user having less bandwidth, the SD outperforms the existing method in some key scenarios. The SD is more robust to strongly LoS channels and clustered users, as it does not rely on spatial multiplexing like other MU methods. Combined with the complexity reduction, this makes the SD an attractive phase selection method.
\end{abstract}


\section{Introduction}

\Glspl{RIS} are the focus of considerable research interest for next generation mobile communications, due to their ability to manipulate the wireless channel. \gls{RIS} can enhance several aspects of system performance, including \gls{SNR}, rank deficiency, and blockage avoidance \cite{wu_towards_2020}. A significant body of work on \gls{RIS} now exists, including design, testbeds, and performance analysis \cite{liu_reconfigurable_2021}. Standards development is underway, with ETSI releasing its first report on \gls{RIS} in 2023 \cite{etsi_reconfigurable_2023}.

Typical \gls{RIS} implementations involve large numbers of elements in an array, where each element can be set to control the phase of the reflected signal. A key challenge is selecting the \gls{RIS} element reflection coefficients to maximize the total \gls{SNR} at the target receiver. Analytically optimal closed-form \gls{RIS} designs for \gls{MU} systems are not tractable, and many existing methods heavily rely on iterative optimization \cite{abeywickrama_intelligent_2020, gao_robust_2024, lee_joint_2024, buzzi_ris_2021,gao_robust_2021,singh_efficient_2022,chen_low_2024}. While these offer excellent performance, they involve significant computational overhead. This becomes particularly challenging when the \gls{RIS} is passive and lacks onboard processing. Moreover, passive \gls{RIS} cannot transmit pilots or process feedback, complicating channel estimation, particularly as the number of users increases. Therefore, reducing both the computational and channel estimation requirements is essential for the practical deployment of \gls{RIS}-assisted \gls{MU} systems.

\subsection{Review of Related Literature}
The design of \gls{RIS} phase selection methods has primarily focused on either iterative optimization methods, or the use of \gls{AI} and \gls{ML} approaches, as analytically optimal closed-form \gls{RIS} designs for \gls{MU} systems are not tractable.

The foundational work in \cite{abeywickrama_intelligent_2020} presents a sum rate maximization method for \gls{MU} \gls{RIS} systems, where the \gls{RIS} phase shift matrix and the \gls{BS} precoding matrix are alternately optimized while considering non-uniform \gls{RIS} reflection amplitudes. A similar problem was solved in \cite{gao_robust_2024} with imperfect \gls{CSI}.

In \cite{lee_joint_2024}, the authors proposed a weighted \gls{DL}–\gls{UL} sum rate optimization framework in which the \gls{DL} and \gls{UL} precoding matrices and the \gls{RIS} phase shift matrix are alternately optimized. For a single-\gls{BS} system, \cite{buzzi_ris_2021} proposes multiple joint optimization strategies over the \gls{RIS} phases and \gls{BS} beamforming vectors to maximize the \gls{SNR}, and for a dual-\gls{BS} system, the authors propose a joint optimization approach over the \gls{RIS} phases and \gls{BS} transmit powers that maximizes the mean \gls{SINR}. 
In \cite{gao_robust_2021}, the transmit power is minimized by alternately optimizing the \gls{RIS} phase shifts (in this case selected from a discrete set) and the \gls{BS} beamforming vector under imperfect \gls{CSI}.

The works in \cite{singh_efficient_2022,chen_low_2024, an_low_2022,xue_multi_2023} focus on lower complexity iterative methods. In \cite{singh_efficient_2022}, a method is proposed that selects the \gls{RIS} phases to increase the sum rate by reducing the total \gls{MSE} of the receiver rather, and in \cite{chen_low_2024}, the precoder, combiner and RIS phases are jointly optimized to minimize the \gls{MSE}. The work in \cite{an_low_2022} proposes a low complexity approach to RIS channel estimation and the optimization of discrete RIS phases to maximize the sum rate, and in \cite{xue_multi_2023}, a low complexity joint optimization method is detailed for a system involving two \glspl{RIS}, where the \gls{RIS} phases and \gls{BS} beamforming are jointly optimized to maximize sum rate.

The application of \gls{AI}/\gls{ML} to the \gls{RIS} phase selection problem has allowed optimal phases to be found for more complex and larger systems. The work in \cite{fondo-ferreiro_neural_2025} trains and tests a fully connected neural network to find the \gls{RIS} phases that minimize the \gls{BER} and maximize the sum rate even with imperfect \gls{CSI}. The authors of \cite{qi_reconfigurable_2025} use deep reinforcement learning to jointly optimize the \gls{RIS} phases and power allocation coefficients to solve a multi-objective optimization problem, where power consumption is minimized and data rate is maximized. In \cite{triwidyastuti_unsupervised_2025}, unsupervised learning is used to jointly optimize the \gls{BS} beamforming vector and the \gls{RIS} phases to maximize the \gls{SNR} in a large-scale \gls{RIS} system with thousands of elements.

All of the methods discussed above are designed to achieve optimal or near-optimal performance, and they deliver excellent results. However, even the low-complexity approaches within these \cite{singh_efficient_2022,chen_low_2024, an_low_2022,xue_multi_2023} rely on iterative procedures and non-trivial matrix operations and optimization steps. As a result, their computational burden can still be significant.

\subsection{Contributions}
In this paper, we take a different direction to the works discussed in the previous section, and target useful performance gains with an extremely low computational complexity  \gls{RIS} phase selection method in a practical \gls{OFDM}-based system. While large-scale systems with multiple \glspl{RIS} are an important topic for research, the availability of exact analytical results for \gls{RIS} systems remains limited. Thus, beginning with a simpler case allows general system insights to be gained through exact expressions and can provide a foundation for analyzing more complex systems in the future.

In \cite{inwood_phase_2023}, we proposed a low-complexity \gls{SD} where a \gls{RIS} is divided into subsurfaces and one subsurface is designed to serve one \gls{UE}. A \gls{MU} system employing \gls{OFDM} is considered. \gls{OFDM} systems are ubiquitous in 4G and 5G systems, and are expected to be so in 6G systems also \cite{3GPP_RAN1_2025}. Each user operates on a different frequency, which enables simple \gls{MF} to be used at the receiver and removes \gls{IUI}. The elements designed for a specific \gls{UE} are selected using the optimal \gls{SU} phase selection method proposed in \cite{singh_optimal_2021}, and are optimal for that \gls{UE} due to the lack of \gls{IUI}. The elements designed for other \glspl{UE} provide uncontrolled scattering for this \gls{UE}. This may assist the \gls{UE}, particularly when there is little natural uncontrolled scattering in the environment. As the elements are designed for a single \gls{UE}, only the channels from that \gls{UE} through one \gls{RIS} subsurface need to be estimated. This leads to a reduction in the channel estimation requirement proportional to the number of \glspl{UE} compared with designs where all UEs operate in the same frequency band. The use of \gls{MF} ensures that the receiver processing is as simple as possible. The \gls{SD} is not intended to compete with the most advanced iterative methods, but rather makes an important contribution to a critical gap in \gls{RIS} research, namely simple low-complexity \gls{MU} phase selection schemes that are viable for real-world implementation.

Motivated by the identified research gaps, we significantly extend the preliminary work in \cite{inwood_phase_2023} by broadening the analysis conducted to a much more general scenario, and proposing multiple extensions to the \gls{SD} that offer further performance gains. The main contributions of this work are as follows:
\begin{itemize}
    \item In our preliminary work in \cite{inwood_phase_2023}, we only considered the performance of the \gls{SD} for a \gls{LoS} \gls{RIS}-\gls{BS} channel and correlated Rayleigh \gls{UE}-\gls{BS} and \gls{UE}-\gls{RIS} channels. In reality, it is likely that all channels will have both \gls{LoS} and \gls{NLoS} components. Therefore, we now significantly extend our work by investigating the performance of this simple and practical method for more general and realistic correlated Ricean fading channels. This versatile analytical tool can be simplified to more specific combinations of \gls{LoS} and \gls{NLoS} channels as desired. 
    \item We also extend the subsurface design to a low-complexity \gls{ISD}. The \gls{SD} assumes that each subsurface is designed independently with no knowledge of phases set for other users. The \gls{ISD} sets the subsurfaces sequentially, utilizing the knowledge of all phases set for other users to improve the design. The \gls{ISD} performs this process once, but it can be performed repeatedly until convergence and the total SNR stops increasing. This is referred to as the \gls{CISD}. The \gls{ISD} and \gls{CISD} offer performance improvements while still benefiting from low computational complexity and a simple \gls{MF} receiver. 
    \item Using analysis and simulations, we investigate the performance of the \gls{SD}, \gls{ISD} and \gls{CISD}. The mean \gls{SNR} of each method is compared for a range of \gls{RIS} dimensions. The mean sum rate supported by the \gls{SD} is compared to a lower-bound benchmark equivalent to no channel knowledge (e.g. randomly selected \gls{RIS} phases) and the \gls{MU} method in \cite{singh_efficient_2022}. This was selected for comparison as it is the lowest complexity existing full \gls{MU} design that we are aware of, and while other methods may offer better performance, it is highly probable that they will require considerably higher system complexity. We thoroughly investigate the difference between the \gls{SD} and \gls{MU} method for a wide range of system parameters, including \gls{RIS} size, correlation between \gls{BS} and \gls{RIS} elements, Ricean K-factor, \gls{UE} location and \gls{UE} clustering, and highlight important scenarios where the \gls{SD} outperforms the \gls{MU} method.
\end{itemize}

\textit{Notation}: Upper and lower boldface letters represent matrices and vectors, respectively. We define vector $\mathbf{v} = [\mathbf{v}_1^T,\dots,\mathbf{v}_K^T]^T$, where $\mathbf{v}_k$ is the $k$-th block of vector $\mathbf{v}$ and $\mathbf{v}_{k,i}$ is the $i$-th element of the $k$-th block. $\mathbf{M}$ is a matrix comprising of submatrices, or blocks. $\mathbf{M}_{r,s,i,j}$ is the $(i,j)$-th element of the  $r$-th block across and $s$-th block down. Matrices are indexed this way except where explicitly defined otherwise. $\mathbf{M}^{(k)}$ denotes a matrix relating to user $k$. $\mathbb{C}$ is the set of complex numbers. $\Re[\cdot]$ is the real operator. $\mathcal{CN}(\boldsymbol\mu,\mathbf{Q})$ represents a complex Gaussian distribution with mean $\boldsymbol\mu$ and covariance matrix $\mathbf{Q}$. $\mathbb{E}[\cdot]$ represents statistical expectation. $(\cdot)^T$, $(\cdot)^*$ and $(\cdot)^\dagger$ are the transpose, conjugate and Hermitian transpose operators, respectively. The angle of a vector, $\mathbf{x}$, of length $N$ is denoted as $\angle\mathbf{x}=[\angle{x}_1,\dots,\angle{x}_N]^T$ and the exponent as $e^{\mathbf{x}}=[e^{{x}_1},\dots,e^{{x}_N}]^T$. $||\cdot||$ is the Euclidean norm. ${}_1F_1(a,b;z)$ is the confluent hypergeometric function and ${}_2F_1(a,b,c;z)$ is the Gaussian hypergeometric function. $L_\nu(\cdot)$ denotes a Laguerre function of non-integer degree $\nu$. $\Gamma(\cdot)$ is the complete gamma function. $f^*(r,\theta)$ is the \gls{PDF} of a complex random variable $X=r\,\mathrm{e}^{j\theta}$, and $f^*(r_1, r_2, \theta_1, \theta_2)$ is the joint  \gls{PDF} of  complex random variables $X_1 = r_1\mathrm{e}^{j\theta_1}$ and $X_2 = r_2\mathrm{e}^{j\theta_2}$. $J_n(\cdot)$ is the $n$-th order Bessel function of the first kind and $I_n(\cdot)$ is the $n$-th order modified Bessel function of the first kind.

\section{System Model}
\begin{figure}[ht]
    \centering
    \includegraphics[trim={0.9cm 0.65cm 0.65cm 0.75cm},clip,scale=0.65]{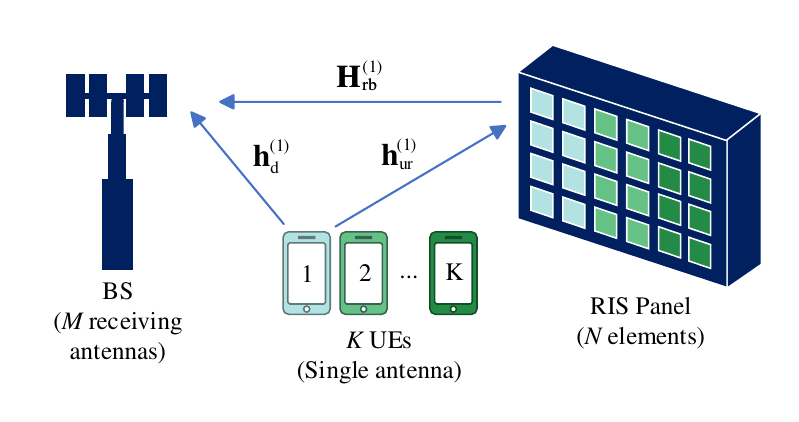}
    \caption{System model showing \gls{UL} channels for \gls{UE} 1 in band 1.}
    \label{fig:system_diagram}
\end{figure}

We consider the \gls{UL} \gls{RIS}-aided system shown in Fig. \ref{fig:system_diagram}, consisting of a \gls{RIS} panel with $N$ elements, a \gls{BS} with $M$ antennas, and $K$ single-antenna users located in the vicinity of the \gls{BS} and \gls{RIS}. A \gls{RIS} link (\gls{UE}-\gls{RIS}-\gls{BS}) and a direct link (\gls{UE}-\gls{BS}) connect the \gls{UE} to the \gls{BS}. The system bandwidth, $B$, is split into $K$ bands of $\frac{B}{K}$ Hz, with one user per band. The phase response of the \gls{RIS}, and therefore the phase shift in the reflected signal, is fixed across the frequency bands. The \gls{RIS} elements are grouped into ``subsurfaces'' and each subsurface is designed for a different user.

For \gls{UE} $k$ in band $k$, let $\mathbf{h}^{(k)}_{\mathrm{d}} \!\in \mathbb{C}^{M\times 1}$, $\mathbf{h}^{(k)}_{\mathrm{ur}} \in \mathbb{C}^{N\times 1}$ and $\mathbf{H}^{(k)}_{\mathrm{rb}} \in \mathbb{C}^{M\times N}$ be the direct, \gls{UE}-\gls{RIS} and \gls{RIS}-\gls{BS} channels, respectively. $\mathbf{\Phi}\in \mathbb{C}^{N\times N}$ is a diagonal matrix of reflection coefficients for the \gls{RIS} which can be given in block diagonal form, where the $k$-th block is designed to enhance the channel for user $k$. $\mathbf{\Phi} = \mathrm{diag}\{\mathbf{\Phi}_1, \dots, \mathbf{\Phi}_K\}$, where $\mathbf{\Phi}_k=\mathrm{diag}(e^{j\phi_{k,1}}, \dots, e^{j\phi_{k,N_k}})$. $N_k=N/K$ is the number of \gls{RIS} elements chosen to support \gls{UE} $k$ and $N=\sum^K_{k=1}N_k$. We consider an equal number of elements per \gls{UE}, and non-equally divided subsurfaces are a potential topic for future work. Elements serving one \gls{UE} are co-located, as this was demonstrated to be the best configuration in \cite{inwood_phase_2023}. The received signal in band $k$ at the \gls{BS} is 
\begin{equation}
    \mathbf{r}_k = (\mathbf{h}^{(k)}_{\mathrm{d}}+\mathbf{H}^{(k)}_{\mathrm{rb}}\mathbf{\Phi h}_{\mathrm{ur}}^{(k)})s_k + \mathbf{n}_k, \label{eq:channel}
\end{equation}
where $s_k$ is the signal being sent from \gls{UE} $k$, $\mathbb{E}[|s_k|^2]=E_s$, and $\mathbf{n}_k\sim \mathbb{C}\mathcal{N}(0,\sigma^2\mathbf{I})$ is \gls{AWGN}. Using the block diagonal form of $\mathbf{\Phi}$, (\ref{eq:channel}) becomes
\begin{multline}    
   \mathbf{r}_k= (\mathbf{h}^{(k)}_{\mathrm{d}} + \mathbf{H}^{(k)}_{\mathrm{rb},k} \mathbf{\Phi}_{k} \mathbf{h}^{(k)}_{\mathrm{ur},k})s_k \\  + \left(\sum_{s\neq k}^{K}\mathbf{H}^{(k)}_{\mathrm{rb},s} \mathbf{\Phi}_{s} \mathbf{h}^{(k)}_{\mathrm{ur},s}\right)s_k + \mathbf{n}_k, \label{eq:channel_split}
\end{multline}
where $\mathbf{H}^{(k)}_{\mathrm{rb}} = \left[\mathbf{H}^{(k)}_{\mathrm{rb},1}, \dots, \mathbf{H}^{(k)}_{\mathrm{rb},K}\right]$, $\mathbf{H}^{(k)}_{\mathrm{rb},i} \in \mathbb{C}^{M\times N_i}$, $\mathbf{h}^{(k)}_{\mathrm{ur}} = \left[\mathbf{h}^{(k)T}_{\mathrm{ur},1}, \dots, \mathbf{h}^{(k)T}_{\mathrm{ur},K}\right]^T$ and $\mathbf{h}^{(k)}_{\mathrm{ur},i} \in \mathbb{C}^{N_i \times 1}$.
\subsection{Channel Model}
\label{sec:channel}
In this work, we consider correlated Ricean channels consisting of a rank-1 \gls{LoS} and correlated Rayleigh component for all links. The \gls{LoS} components are comprised of steering vectors and the scattered components adopt the Kronecker correlation model. Therefore, the channels for \gls{UE} $k$ are
\begin{align}
    \mathbf{h}^{(k)}_{\mathrm{d}} & =\sqrt{\beta^{(k)}_\mathrm{d}}\left(\eta^{(k)}_\mathrm{d}\Tilde{\mathbf{h}}^{(k,\mathrm{LoS})}_{\mathrm{d}}+ \zeta^{(k)}_\mathrm{d}\Tilde{\mathbf{h}}_{\mathrm{d}}^{(k,\mathrm{SC})}\right), \label{eq:hdexpand} \\
    \mathbf{H}^{(k)}_{\mathrm{rb}} & = \sqrt{\beta^{(k)}_\mathrm{rb}}\left(\eta^{(k)}_\mathrm{rb}\Tilde{\mathbf{H}}_{\mathrm{rb}}^{(k,\mathrm{LoS})} + \zeta^{(k)}_\mathrm{rb}\Tilde{\mathbf{H}}^{(k,\mathrm{SC})}_{\mathrm{rb}}\right),\label{eq:Hrbexpand}\\
    \mathbf{h}^{(k)}_{\mathrm{ur}} & = \sqrt{\beta^{(k)}_\mathrm{ur}}\left(\eta^{(k)}_\mathrm{ur}\Tilde{\mathbf{h}}_{\mathrm{ur}}^{(k,\mathrm{LoS})}+ \zeta^{(k)}_\mathrm{ur}\Tilde{\mathbf{h}}_{\mathrm{ur}}^{(k,\mathrm{SC})}\right), \label{eq:hruexpand}
\end{align}
with
\begin{align}
	\eta^{(k)}_\mathrm{d}=\sqrt{\frac{\kappa^{(k)}_\mathrm{d}}{\kappa^{(k)}_\mathrm{d}+1}},\quad \Label{eq:etad} & \enskip\! & \zeta^{(k)}_\mathrm{d}=\sqrt{\frac{1}{\kappa^{(k)}_\mathrm{d}+1}}, \quad \Label{eq:zetad} \notag \\
	\eta^{(k)}_\mathrm{rb}=\sqrt{\frac{\kappa^{(k)}_\mathrm{rb}}{\kappa^{(k)}_\mathrm{rb}+1}}, \quad \Label{eq:etarb} &\enskip\! & \zeta^{(k)}_\mathrm{rb}=\sqrt{\frac{1}{\kappa^{(k)}_\mathrm{rb}+1}}, \quad \Label{eq:zetarb} \notag \\
	\eta^{(k)}_\mathrm{ur}=\sqrt{\frac{\kappa^{(k)}_\mathrm{ur}}{\kappa^{(k)}_\mathrm{ur}+1}}, \quad \Label{eq:etaur} &\enskip\! & \zeta^{(k)}_\mathrm{ur}=\sqrt{\frac{1}{\kappa^{(k)}_\mathrm{ur}+1}}, \quad \Label{eq:zetaur} \notag
\end{align}
where $\beta^{(k)}_\mathrm{d}$, $\beta^{(k)}_\mathrm{rb}$ and $\beta^{(k)}_\mathrm{ur}$ are the channel gains, $\kappa^{(k)}_\mathrm{d}$, $\kappa^{(k)}_\mathrm{rb}$ and $\kappa^{(k)}_\mathrm{ur}$ are the Ricean K-factors, $\Tilde{\mathbf{h}}_{\mathrm{d}}^{(k,\mathrm{LoS})}$, $\Tilde{\mathbf{H}}_{\mathrm{rb}}^{(k,\mathrm{LoS})}$ and $\Tilde{\mathbf{h}}_{\mathrm{ur}}^{(k,\mathrm{LoS})}$ are the \gls{LoS} channel components for the $k$-th \gls{UE} and $\Tilde{\mathbf{h}}_{\mathrm{d}}^{(k,\mathrm{SC})}$, $\Tilde{\mathbf{H}}_{\mathrm{rb}}^{(k,\mathrm{SC})}$ and $\Tilde{\mathbf{h}}_{\mathrm{ur}}^{(k,\mathrm{SC})}$ are the \gls{SC} components for the $k$-th \gls{UE}. Here,
\begin{alignat}{4}
    \Tilde{\mathbf{h}}_{\mathrm{d}}^{(k,\mathrm{LoS})} &= \mathbf{a}^{(k)}_\mathrm{d},& \qquad\Tilde{\mathbf{h}}_{\mathrm{d}}^{(k,\mathrm{SC})} &= \mathbf{R}^{(k)1/2}_\mathrm{d}\mathbf{u}^{(k)}_{\mathrm{d}},\notag\\
    \Tilde{\mathbf{H}}_{\mathrm{rb}}^{(k,\mathrm{LoS})} &= \mathbf{a}^{(k)}_\mathrm{b}\mathbf{a}^{(k)\dagger}_\mathrm{r}, & \qquad\Tilde{\mathbf{H}}_{\mathrm{rb}}^{(k,\mathrm{SC})} &= \mathbf{R}_{\mathrm{b}}^{(k)1/2}\mathbf{U}_{\mathrm{rb}}^{(k)}\mathbf{R}_{\mathrm{r}}^{(k)1/2},\notag\\
    \Tilde{\mathbf{h}}_{\mathrm{ur}}^{(k,\mathrm{LoS})} &= \mathbf{a}_\mathrm{ur}^{(k)},& \qquad\Tilde{\mathbf{h}}_{\mathrm{ur}}^{(k,\mathrm{SC})} &= \mathbf{R}_{\mathrm{ur}}^{(k){1/2}}\mathbf{u}^{(k)}_{\mathrm{ur}},\notag
\end{alignat}
where $\mathbf{a}^{(k)}_\mathrm{d}$ is the steering vector of the \gls{LoS} ray for the direct link at the \gls{BS}, $\mathbf{a}^{(k)}_\mathrm{b}$ and $\mathbf{a}^{(k)}_\mathrm{r}$ are steering vectors for the \gls{LoS} ray for the \gls{RIS}-\gls{BS} link at the \gls{BS} and \gls{RIS}, respectively, and $\mathbf{a}^{(k)}_\mathrm{ur}$ is the steering vector of the \gls{LoS} ray for the \gls{UE}-\gls{RIS} link at the \gls{RIS}. $\mathbf{R}^{(k)}_{\mathrm{b}}$, $\mathbf{R}^{(k)}_{\mathrm{r}}$, $\mathbf{R}^{(k)}_{\mathrm{d}}$ and $\mathbf{R}^{(k)}_{\mathrm{ur}}$ are the correlation matrices for the \gls{RIS}-\gls{BS} link at the \gls{BS} and \gls{RIS} ends, the \gls{UE}-\gls{BS} link and \gls{UE}-\gls{RIS} link, respectively, and $\mathbf{u}^{(k)}_{\mathrm{d}}, \mathbf{U}^{(k)}_{\mathrm{rb}}$ and $\mathbf{u}^{(k)}_{\mathrm{ur}}$ are all matrices and vectors containing independent and identically distributed $\mathbb{C}\mathcal{N}(0,1)$ entries.

This work focuses on strongly \gls{LoS} $\mathbf{H}_\mathrm{rb}$ channels, where one ray is dominant but a scattered component is also present. This is a common and reasonable assumption for \gls{RIS} systems, as the \gls{RIS} can be located to exploit the \gls{LoS} channel as seen in \cite{wu_intelligent_2019, nadeem_asymptotic_2020, alouini_exploiting_2021}. We also assume that reliable \gls{CSI} is available for each individual link. Prior work has demonstrated that the \gls{UE}-\gls{RIS} and \gls{RIS}-\gls{BS} channels can be estimated separately, as in \cite{hu_two_2021, ling_two_2023}. Hence, we anticipate that analogous methods could be applied here.

\section{Phase Selection Methods}
\label{sec:phase_selection_methods}

Partitioning the $N$ \gls{RIS} elements into $K$ subsurfaces each designed independently for a specific \gls{UE} enables low-complexity and efficient \gls{RIS} phase selection. This section outlines three methods that utilise this concept: a simple approach that uses the \gls{SU} optimal phase selection method for when $\mathbf{H}_\mathrm{rb}$ is rank-1, a sub-optimal extension of this for when $\mathbf{H}_\mathrm{rb}$ has a scattered component, and a higher-performing but more complex iterative version.

\subsection{LoS Phase Selection Method}
\label{sec:LOS_phase_selection}
As we first proposed in \cite{inwood_phase_2023}, the phases of a subsurface designed for the $k$-th \gls{UE} can be set according to the \gls{SU} method set out in \cite{singh_optimal_2021}, so that
\begin{equation}
	\label{eq:Phi}
	\mathbf{\Phi}_k = \nu_k \, \mathrm{diag}\left(e^{j\left(\angle \mathbf{a}^{(k)}_{\mathrm{r},k}- \angle \mathbf{h}^{(k)}_{\mathrm{ur},k}\right)}\right),
\end{equation}
where
\begin{equation}
    \label{eq:nus}
    \nu_k =\frac{\mathbf{a}^{(k)\dagger}_\mathrm{b}\mathbf{h}^{(k)}_{\mathrm{d}}}{|\mathbf{a}^{(k)\dagger}_\mathrm{b}\mathbf{h}^{(k)}_{\mathrm{d}}|}.
\end{equation}
The phases set for each \gls{UE} are optimal for that \gls{UE} when $\mathbf{H}_\mathrm{rb}^{(k)}$ is purely \gls{LoS}, and close to optimal when $\mathbf{H}_\mathrm{rb}^{(k)}$ has a strong \gls{LoS} component. This method leads to a $K$-fold reduction in the required \gls{CSI} compared to typical \gls{RIS} designs, due to only requiring the estimation of $N_k$ rather than $N$ channel elements per user. There are no complex matrix operations required and a simple \gls{MF} receiver can be used. These advantages make this approach a very attractive option for implementation.

\subsection{Extended Subsurface Design for NLoS Systems}
\label{sec:NLoS_phase_selection}
The design in (\ref{eq:Phi}) and (\ref{eq:nus}) works well for strongly \gls{LoS} channels. However, if there is a strong scattered component, a different approach provides better results. As there is no optimal closed form \gls{SU} phase selection method when $\mathbf{H}^{(k)}_\mathrm{rb}$ is not rank-1, we propose the following adaptation to the method outlined in Sec. \ref{sec:LOS_phase_selection}, referred to as the \gls{ESD}. As we proposed in \cite{inwood_phase_2023}, we take the \gls{SVD} of $\mathbf{H}^{(k)}_{\mathrm{rb},k}$, so that $\mathbf{H}^{(k)}_{\mathrm{rb},k} = \mathbf{UDV^\dagger}$, where $\mathbf{D} = \mathrm{diag}(D_{11}, D_{22}, \dots, D_{mm})$, $m = \min(M, N_k)$ and the singular values are $D_{11} \geq D_{22} \geq ... \geq D_{mm}$. The strongest rank-1 component of $\mathbf{H}^{(k)}_{\mathrm{rb},k}$ is $D_{11}\mathbf{u}_k\mathbf{v}_k^\dagger$, where $\mathbf{u}_k$ and $\mathbf{v}_k$ are the leading left and right singular vectors, respectively. This has the same structure as a pure \gls{LoS} channel, so the same design can be used by considering just the strongest component, giving,
\begin{equation}
    \label{eq:Phi_NLoS}
    \mathbf{\Phi}_\mathrm{k} = \omega_k\,\mathrm{diag}\left(e^{j\left(\mathbf{\angle v}_{k}-\mathbf{\angle h}^{(k)}_{\mathrm{ur},k}\right)}\right),
\end{equation}
where 
\begin{equation}
    \omega_k=\frac{|\mathbf{h}_{\mathrm{ru},k}^{(k)}|^T \mathrm{diag}(e^{-j\angle \mathbf{v}_k}) \mathbf{H}_{\mathrm{br},k}^{(k)\dagger} \mathbf{h}_{\mathrm{d}}^{(k)}}{\left||\mathbf{h}_{\mathrm{ru},k}^{(k)}|^T \mathrm{diag}(e^{-j\angle \mathbf{v}_k}) \mathbf{H}_{\mathrm{br},k}^{(k)\dagger}\mathbf{h}_{\mathrm{d}}^{(k)}\right|}.
\end{equation}

\subsection{Iterative Phase Alignment}
\label{sec:ISD}

The subsurface methods outlined in Secs. \ref{sec:LOS_phase_selection} and \ref{sec:NLoS_phase_selection} assume that each subsurface is designed independently, with no knowledge of phases that have been set for other users. Since the reflections produced by the other subsurfaces manifest as random scattering for a given user, setting the subsurfaces sequentially allows this randomness to be incorporated into the design and used to inform the configuration of the remaining subsurfaces.
 
The methods in the previous sections select phases to maximize the norm of the subsurface's \gls{RIS} link, $\mathbf{H}^{(k)}_{\mathrm{rb},k} \mathbf{\Phi}_{k} \mathbf{h}^{(k)}_{\mathrm{ur},k}$, and then rotate these phases to align with the non-\gls{RIS}-controlled direct link, $\mathbf{h}^{(k)}_{\mathrm{d}}$. The \gls{ISD} instead involves selecting phases to maximize the \gls{RIS} link, and then aligning these with a combination of the direct path and the \gls{RIS} paths that have already been set for other \glspl{UE}. This is done by including this information into $\nu_k$ (\gls{LoS} case) or $\omega_k$ (\gls{NLoS} case). The subsurfaces are set from the weakest \gls{UE}-\gls{RIS} channel to the strongest, so that the most impactful channels can benefit from the most information. This process is detailed in Algorithm \ref{alg:ISD}, where index$(\cdot)$ refers to a function that returns the index of elements in a list and sort$(\cdot)$ refers to a function that reorders elements of a list from smallest to largest.

This iterative method improves the mean \gls{SNR} as shown in Sec. \ref{sec:numres_sub} at the cost of a significant increase in \gls{CSI} requirements at the \gls{BS}. $\mathbf{h}_\mathrm{ur}^{(k)}$ contains $N$ channel elements, with $\frac{N}{K}$ corresponding to those impacted by a specific subsurface. The iterative method requires the computation of the magnitude of a user's \gls{UE}-\gls{RIS} link to rank channels based on their impact, so knowledge of the amplitude of all $N$ channel elements is required. In contrast, the methods in Secs. \ref{sec:LOS_phase_selection} and \ref{sec:NLoS_phase_selection} only consider the impact on a \gls{UE} from each subsurface, so only the subsurface specific $\frac{N}{K}$ elements are required. The benefits of the iterative method vary with channel conditions such as correlation and link power, as will be shown in Sec. \ref{sec:numres_sub}.

\begin{algorithm}
\caption{Iterative Phase Alignment Method}
\begin{algorithmic} 
\label{alg:ISD}
\STATE $\mathit{user\_order} =\mathrm{index(sort(}||\mathbf{h}^{(1)}_\mathrm{ur}||^2, ... , ||\mathbf{h}^{(K)}_\mathrm{ur}||^2))$
\FOR{$k = 1$ to $K$}
    \STATE $\mathit{index\_k} = \mathit{user\_order}(k)$
    \STATE $\mathit{fixed} = 0$
    \FOR{$l = 1$ to $k-1$}
    \STATE $\mathit{index\_l} = \mathit{user\_order}(l)$
    \STATE $\mathit{fixed} = \mathit{fixed} + \mathbf{H}^{(\mathit{index\_k})}_{\mathrm{rb},\mathit{index\_l}} \mathbf{\Phi}_{\mathit{index\_l}}\mathbf{h}^{(\mathit{index\_k})}_{\mathrm{ur},\mathit{index\_l}}$
    \ENDFOR
    \STATE $\mathit{direct} = \mathbf{h}^{(\mathit{index\_k})}_{\mathrm{d}} + \mathit{fixed}$    
    \STATE $\nu_\mathit{index\_k} = \dfrac{\mathbf{a}_\mathrm{b}^{(\mathit{index\_k})\dagger}\times\mathit{direct}}{|\mathbf{a}_\mathrm{b}^{(\mathit{index\_k})\dagger}\times\mathit{direct}|}$
    \STATE $\mathbf{\Phi}_\mathrm{\mathit{index\_k}} = \nu_{\mathit{index\_k}}\,\mathrm{diag}\left(\mathrm{e}^{j\left(\angle\mathbf{a}^{(\mathit{index\_k})}_{\mathrm{r}} - \angle\mathbf{h}^{(\mathit{index\_k})}_{\mathrm{ur},\mathit{index\_l}}\right)}\right)$
\ENDFOR
\end{algorithmic}
\end{algorithm}

\section{Analysis}
In this section, we focus on an analysis of the mean \gls{SNR} as prior work \cite{singh_optimal_2021,inwood_phase_2023,shaikh_on_2022} has shown it to be tractable and to deliver system insights. The accuracy of this rate bound was investigated for the \gls{SD} in our work in \cite{inwood_phase_2023}, and it was found to be tight for all values. In particular, we derive the mean \gls{SNR} for a correlated Ricean environment, where $\mathbf{h}_\mathrm{d}$, $\mathbf{H}_\mathrm{rb}$ and $\mathbf{h}_\mathrm{ur}$ all consist of a rank-1 \gls{LoS} and a scattered component. We consider the important case where $\mathbf{H}_\mathrm{rb}$ has a strong, but not pure, \gls{LoS} component. This leads to a stable, slow-moving \gls{RIS}-\gls{BS} channel, allowing the \gls{SD} to be used with close to optimal performance for a given subsurface \cite{singh_optimal_2022}. Unfortunately, as far as we know, a closed-form result for the mean \gls{SNR} is not available using the \gls{ESD}. However, the SD remains highly accurate in the considered scenario. This enables the derivation of closed-form results, which are particularly valuable for understanding how system behavior arises from underlying mathematical structure. Therefore, $\mathbf{\Phi}$ in \eqref{eq:Phi} is used in this section. This general case is then simplified to two special cases where $\mathbf{h}_\mathrm{d}$ and $\mathbf{h}_\mathrm{ur}$ are correlated Rayleigh channels; in the first $\mathbf{H}_\mathrm{rb}$ is correlated Ricean, and in the second it is a rank-1 \gls{LoS} channel. The received signal $\mathbf{r}_k$ in (\ref{eq:channel_split}) can be rewritten as
\begin{equation}
\label{eq:rk}
	\mathbf{r}_k = \left(\mathbf{h}^{(k)}_{\mathrm{d}} + \mathbf{f}_k + \mathbf{g}_k\right)s_k + \mathbf{n}_k, 
\end{equation}
where $\mathbf{f}_k = \mathbf{H}^{(k)}_{\mathrm{rb},k}\mathbf{\Phi}_k \mathbf{h}^{(k)}_{\mathrm{ur},k}$ and 
$\mathbf{g}_k = \sum_{s\neq k}^{K}\mathbf{H}^{(k)}_{\mathrm{rb},s}\mathbf{\Phi}_s \mathbf{h}^{(k)}_{\mathrm{ur},s}$. Thus, the \gls{SNR} is given by 
\begin{multline}
    \mathrm{SNR}_k =  \frac{E_s}{\sigma^2} \Big[\mathbf{h}^{(k)\dagger}_{\mathrm{d}} \mathbf{h}^{(k)}_{\mathrm{d}} + 2\Re(\mathbf{h}_{\mathrm{d}}^{(k)\dagger} \mathbf{f}_k) + 2\Re(\mathbf{h}_{\mathrm{d}}^{(k)\dagger} \mathbf{g}_k) \\ + 2\Re(\mathbf{f}_k^\dagger \mathbf{g}_k) + \mathbf{f}_k^\dagger \mathbf{f}_k + \mathbf{g}_k^\dagger \mathbf{g}_k\Big]. \label{eq:SNRterms}
\end{multline}
\subsection{Mean SNR for Correlated Ricean Systems}
\label{sec:ESNR_Ricean}
We now consider the mean value of each term in (\ref{eq:SNRterms}) to calculate $\mathbb{E}[\mathrm{SNR}_k]$ for correlated Ricean systems. From \cite{singh_optimal_2022},
\begin{equation}
    \mathbb{E}[\mathbf{h}^{(k)\dagger}_{\mathrm{d}} \mathbf{h}^{(k)}_{\mathrm{d}}] = M\beta_\mathrm{d}^{(k)},  \label{eq:Ehdhd}
\end{equation}
and
\begin{multline}
    \mathbb{E}[\mathbf{h}^{(k)\dagger}_{\mathrm{d}}\mathbf{f}_\mathrm{k}] = \frac{N_k\pi||\mathbf{R}^{(k)1/2}_{\mathrm{d}}\mathbf{a}^{(k)}_\mathrm{b}||\sqrt{\!\beta^{(k)}_{\mathrm{d}}\!\beta^{(k)}_{\mathrm{rb}}\!\beta^{(k)}_{\mathrm{ur}}}}{4} \\ 
    \times \eta_\mathrm{rb}^{(k)}\zeta_\mathrm{d}^{(k)}\zeta_\mathrm{ur}^{(k)} L_{1/2}\!\left(\!\!-\kappa_\mathrm{ru}^{(k)}\!\right)\!L_{1/2}\!\!\left(\frac{\!\!-\kappa_\mathrm{d}^{(k)}|\mathbf{a}_\mathrm{b}^{(k)\dagger}\mathbf{a}_\mathrm{d}^{(k)}|^2}{||\mathbf{R}^{(k)1/2}_{\mathrm{d}}\mathbf{a}_\mathrm{b}^{(k)}||^2}\!\right)\!\!.\!\!\!\label{eq:Ehdf}
\end{multline}
\begin{remark}
    $\mathbb{E}[\mathbf{h}_\mathrm{d}^{(k)\dagger}\mathbf{h}^{(k)}_\mathrm{d}]$ represents the gain of the direct channel, and $\mathbb{E}[\mathbf{h}_\mathrm{d}^{(k)\dagger}\mathbf{f}_\mathrm{k}]$ is the cross product of the direct channel and channel through the subsurface designed for \gls{UE} $k$. Higher element spacing at the BS increases the \gls{SNR}, due to the $||\mathbf{R}_\mathrm{d}\mathbf{a}_\mathrm{b}||$ component of (\ref{eq:Ehdf}). This term reduces if \gls{BS} elements are placed closely together, as in \cite{singh_optimal_2021}.
\end{remark}
All remaining terms are derived in Appendix A and their results are listed below.\\
\begin{multline}
    \mathbb{E}[\mathbf{h}^{(k)\dagger}_{\mathrm{d}}\mathbf{g}_\mathrm{k}] = \sqrt{\beta^{(k)}_{\mathrm{d}}\beta^{(k)}_{\mathrm{rb}}\beta^{(k)}_{\mathrm{ur}}}\eta^{(k)}_{\mathrm{d}}\eta^{(k)}_{\mathrm{rb}}\eta^{(k)}_{\mathrm{ur}} \mathbf{a}^{(k)\dagger}_\mathrm{d}\mathbf{a}_\mathrm{b}^{(k)} \\ \times\sum_{s\neq k}\mathbf{a}_{\mathrm{r},s}^{(k)\dagger}\mathbf{C}(s)\mathbf{a}_{\mathrm{ur},s}^{(k)}, \label{eq:Ehdgk}
\end{multline}
where
\begin{multline}
    \mathbf{C}(x)\!=\!\frac{\pi\sqrt{\!\kappa_\mathrm{d}^{(x)}\!\kappa_\mathrm{ur}^{(x)}}\mathbf{a}_\mathrm{b}^{(x)\dagger}\mathbf{a}_\mathrm{d}^{(x)}\!}{4\sqrt{\mathbf{a}_\mathrm{b}^{(x)\dagger}\mathbf{R}_\mathrm{d}^{(x)}\mathbf{a}_\mathrm{b}^{(x)}}}\mathrm{diag}\!\left(\!\mathrm{e}^{j\left(\angle\mathbf{a}_{\mathrm{r},x}^{(x)} - \angle\mathbf{a}_{\mathrm{ur},x}^{(x)}\right)}\!\right) \\ \times{}_1F_1\!\!\left(\!\frac{1}{2},2,-\frac{|\mathbf{a}^{(x)\dagger}_\mathrm{b}\mathbf{a}^{(x)}_\mathrm{d}|^2\kappa^{(x)}_\mathrm{d}}{\mathbf{a}_\mathrm{b}^{(x)\dagger}\mathbf{R}_\mathrm{d}^{(x)}\mathbf{a}_\mathrm{b}^{(x)}}\!\right)\!{}_1F_1\!\!\left(\!\frac{1}{2},2,-\kappa_\mathrm{ur}^{(x)}\!\right). \label{eq:Cx}
\end{multline}
\begin{remark}
    \label{rem:Ehdgk}
    $\mathbb{E}[\mathbf{h}_\mathrm{d}^{(k)\dagger}\mathbf{g}_\mathrm{k}]$ represents the cross products of the direct channel and the channels through subsurfaces designed for other \glspl{UE}. The complexity of the summation term in \eqref{eq:Ehdgk} makes it difficult to interpret in detail, but the size of the term is increased, as expected, by alignment of the direct and \gls{RIS} paths. Note that the summation of a large number of phases in $\mathbf{a}^{(k)}_{\mathrm{r},s}$ and $\mathbf{a}^{(k)}_{\mathrm{ur},s}$ is likely to cause significant cancellation, so this is not a dominant term.
\end{remark}
\begin{align}
    &\mathbb{E}[\mathbf{f}_\mathrm{k}^\dagger\mathbf{g}_\mathrm{k}] = \frac{M\pi\beta_\mathrm{rb}^{(k)}\beta_\mathrm{ur}^{(k)}\sqrt{\kappa_{\mathrm{d}}^{(k)}\kappa_{\mathrm{ur}}^{(k)}}\mathbf{a}_\mathrm{d}^{(k)\dagger}\mathbf{a}_\mathrm{b}^{(k)}\mathrm{e}^{-\kappa_\mathrm{ur}^{(k)}}}{4\!\left(1\!+\!\kappa_\mathrm{rb}^{(k)}\right)\!\!\left(1\!+\!\kappa_\mathrm{ur}^{(k)}\right)\!\sqrt{\mathbf{a}_\mathrm{b}^{(k)\dagger}\mathbf{R}_\mathrm{d}^{(k)}\mathbf{a}_\mathrm{b}^{(k)}}}\notag \\ & \times\!{}_1F_1\!\left(\!\frac{1}{2},2,-\frac{|\mathbf{a}_\mathrm{b}^{(k)\dagger}\mathbf{a}_\mathrm{d}^{(k)}|^2\kappa_\mathrm{d}^{(k)}}{\mathbf{a}_\mathrm{b}^{(k)\dagger}\mathbf{R}_\mathrm{d}^{(k)}\mathbf{a}_\mathrm{b}^{(k)}}\!\right)\!\sum_{s\neq k}\sum_{i=1}^{N_k}\sum_{j=1}^{N_k}\bigg[\!\Big(\mathbf{a}^{(k)}_{\mathrm{ur},s,j}\!-\! \rho^{(k)}_{ks} \notag \\ & \times\mathbf{a}_{\mathrm{ur},k,i}^{(k)}\Big){}_1F_1\!\!\left(\!\frac{3}{2},1,\kappa_\mathrm{ur}^{(k)}\!\right)\!+\!\frac{3}{2}\rho^{(k)}_{ks}{}_1F_1\!\!\left(\!\frac{5}{2},2,\kappa_\mathrm{ur}^{(k)}\!\right)\!\mathrm{e}^{j\angle\mathbf{a}_{\mathrm{ur},k,i}^{(k)}}\!\bigg] \notag \\ & \times\!\left(\!\kappa_\mathrm{rb}^{(k)}\mathbf{a}^{(k)}_{\mathrm{r},k,i}\mathbf{a}^{(k)*}_{\mathrm{r},s,j}\!+\! r_{ks}^{(k)}\right)\!(\mathbf{C}(s))_{j,j}\,\mathrm{e}^{-j\angle\mathbf{a}_{\mathrm{r},k,i}^{(k)}}, \label{eq:Efkgk}
\end{align}
where $\mathbf{C}(x)$ is given in (\ref{eq:Cx}), and, dropping subscripts for readability, $\rho^{(k)}_{ks}=\mathbf{R}^{(k)}_{\mathrm{ur},k,s,i,j}$ and $r^{(k)}_{ks}=\mathbf{R}^{(k)}_{\mathrm{r},k,s,i,j}$.
\begin{remark}
    $\mathbb{E}[\mathbf{f}_\mathrm{k}^{\dagger}\mathbf{g}_\mathrm{k}]$ represents the cross products of the channel through the subsurface designed for the \gls{UE} and the channels through subsurfaces designed for other \glspl{UE}. As in Remark \ref{rem:Ehdgk}, the main feature of \eqref{eq:Efkgk} is that the summation of large numbers of weighted phases tends to cause considerable cancellation, reducing the importance of this term.
\end{remark}
\begin{multline}
    \mathbb{E}[\mathbf{f}_\mathrm{k}^\dagger\mathbf{f}_\mathrm{k}] = M\beta_\mathrm{rb}^{(k)}\beta_\mathrm{ur}^{(k)}\bigg[N_k\left(\eta_\mathrm{rb}^{(k)2}+\zeta_\mathrm{rb}^{(k)2}\right) \\ + \sum_{i=1}^{N_k}\sum_{j\neq i}\left(\eta_\mathrm{rb}^{(k)2}+\zeta_\mathrm{rb}^{(k)2}\mathbf{A}_{i,j}\right)F_R\bigg], \label{eq:Efkfk}
\end{multline}
where 
\begin{equation}
        \mathbf{A} = \mathrm{diag}\left(\mathbf{a}_\mathrm{r}^{(k)\dagger}\right)\mathbf{R}_{\mathrm{r},k,k}^{(k)}\mathrm{diag}\left(\mathbf{a}_\mathrm{r}^{(k)}\right),
\end{equation}
\begin{align}
    &F_R = \frac{(1-|\rho^{(k)}_{kk}|^2)^2}{1+\kappa_\mathrm{ur}^{(k)}}\!\exp{\!\!\left(\!-\frac{2\kappa_\mathrm{ru}^{(k)}(1-\mu_{cf})}{1-|\rho^{(k)}_{kk}|^2}\!\right)}\!\sum_{m=0}^\infty\sum_{n=0}^m \notag \\
    &\times\!\cos(n\phi_f)\frac{\epsilon_n|\rho^{(k)}_{kk}|^{2m-n}}{m!(m-n)!(n!)^2}\!\left(\!\frac{\kappa_\mathrm{ru}^{(k)}(1+|\rho^{(k)}_{kk}|^2-2\mu_{cf}))}{1-|\rho^{(k)}_{kk}|^2}\!\!\right)^{\!\!n}\!\!\ \notag \\
    &\times\!\Gamma^2\!\!\left(\!\!m\!+\!\frac{3}{2}\!\right)\!{}_1F_1^2\!\!\left(\!m\!+\!\frac{3}{2}, n\!+\!1, \frac{\kappa_\mathrm{ru}^{(k)}(1+|\rho^{(k)}_{kk}|^2-2\mu_{cf}))}{1-|\rho^{(k)}_{kk}|^2}\!\right)\!,
\end{align}
$\rho_{kk}^{(k)}=\mathbf{R}^{(k)}_{\mathrm{ur},k,k,i,j}$, $\phi_f=\angle((1+|\rho_{kk}^{(k)}|^2)\mu_{cf}\kappa_\mathrm{ur}^{(k)} -2\kappa_\mathrm{ur}^{(k)}|\rho_{kk}^{(k)}|^2+j(1-|\rho_{kk}^{(k)}|^2)\mu_{sf}\kappa_\mathrm{ur}^{(k)})$,  $\mu_{cf}\!=\!\rho^{(k)}_{kk}\!\cos\!\left(\!\angle\mathbf{a}_{\mathrm{ur},k,i}^{(k)}\!-\!\angle\mathbf{a}_{\mathrm{ur},k,j}^{(k)}\!\right)$ and $\mu_{sf}\!=\!\rho^{(k)}_{kk}\!\sin\!\left(\!\angle\mathbf{a}_{\mathrm{ur},k,i}^{(k)}\!-\!\angle\mathbf{a}_{\mathrm{ur},k,j}^{(k)}\!\right)$.
\begin{remark}
    $\mathbb{E}[\mathbf{f}_\mathrm{k}^\dagger\mathbf{f}_\mathrm{k}]$ represents the gain of the channel through the subsurface designed for \gls{UE} $k$. It includes Gaussian hypergeometric functions, which increase monotonically with the final argument. The final argument increases with the spatial correlation between subsurface elements, and this increase occurs when subsurface elements are placed closer together. Hence, placing the subsurface elements close together actually \emph{increases} the mean \gls{SNR}. 
\end{remark}
\begin{align}
    &\mathbb{E}[\mathbf{g}_\mathrm{k}^\dagger\mathbf{g}_\mathrm{k}] = \frac{M\beta_\mathrm{rb}^{(k)}\beta_\mathrm{ur}^{(k)}}{\left(\kappa_\mathrm{rb}^{(k)} + 1\right)\!\!\left(\kappa_\mathrm{ur}^{(k)} + 1\right)}\Bigg[\sum_{s=t\neq k}\sum_{i=1}^{N_k}\sum_{j=1}^{N_k}\,G_R\notag \\
    &\quad\times\mathrm{e}^{-j\angle\mathbf{a}_{\mathrm{r},s,i}^{(s)}}\mathrm{e}^{j\angle\mathbf{a}_{\mathrm{r},s,j}^{(s)}}\Big(\kappa_\mathrm{ur}^{(k)}  \mathbf{a}_{\mathrm{ur},s,i}^{(k)*}\mathbf{a}_{\mathrm{ur},s,j}^{(k)} + \rho^{(k)}_{ss}\Big)\Big(\kappa_\mathrm{rb}^{(k)} \notag \\ &\quad \times\mathbf{a}_{\mathrm{r},s,i}^{(k)}\mathbf{a}_{\mathrm{r},s,j}^{(k)*} + r^{(k)}_{ss}\Big)\!+\!\!\sum_{s\neq t\neq k}\sum_{i=1}^{N_k}\sum_{j=1}^{N_k}(\mathbf{C}^*\!(s)\!)_{i,i}(\mathbf{C}(t)\!)_{j,j}\notag \\ &\quad\times\!\Big(\!\kappa_\mathrm{ur}^{(k)}
    \mathbf{a}_{\mathrm{ur},s,i}^{(k)*}\mathbf{a}_{\mathrm{ur},t,j}^{(k)}\!+\! \rho^{(k)}_{st}\!\Big)\!\Big(\!\kappa_\mathrm{rb}^{(k)}\mathbf{a}_{\mathrm{r},s,i}^{(k)} \mathbf{a}_{\mathrm{r},t,j}^{(k)*}\!+\!r^{(k)}_{st}\!\Big)\!\Bigg]\!, \label{eq:Egkgk}
\end{align}
where
\begin{align}
        &G_R = \!\frac{2\!\left(\!1\!+\!\kappa_\mathrm{ur}^{(s)}\!\right)^{\!\!2}\!\mathrm{e}^{j\!\left(\!\angle(\!1-\mu_c+j\mu_s\!)+\!\angle\mathbf{a}^{(s)}_{\mathrm{ur},s,j}\!\right)}\!}{\!\left(\!1\!-\!\rho_{ss}^{(s)2}\!\right)\!\mathrm{e}^{j\!\left(\!\angle(\!1-\mu_c-j\mu_s\!)+\!\angle\mathbf{a}^{(s)}_{\mathrm{ur},s,i}\!\right)}\!}\!\exp\!\!\left(\!\!\frac{-2\kappa_\mathrm{ur}^{(s)}\!(\!1\!-\!\mu_c)}{1\!-\!\rho_{ss}^{(s)2}}\!\!\right)\! \notag \\ &\times \sum_{n=0}^\infty\sum_{m=0}^\infty\frac{\epsilon_n\kappa_0^{2m+n}}{2^{m+n}m!\Gamma(m\!+\!n\!+\!1)}\left(\mathcal{F}(3)\mathrm{e}^{jn\phi}\!+\! \mathcal{F}(1)\mathrm{e}^{-jn\phi}\right)\!, \label{eq:GR_body}
\end{align}
\begin{align}
    &\mathcal{F}(x)\!=\!\frac{\Gamma^2\!\left(n\!+\!m\!+\!\frac{x}{2}\right)\!\!\left(\!\zeta\sqrt{\kappa_\mathrm{ur}^{(s)}(1\!+\!|\rho_{ss}^{(s)}|^2\!-\!2\mu_c}\right)^{\!\!\!2n-4+2x}}{4^{n-1+x}\,\Gamma^2(n-1+x)\left(\frac{1+\kappa^{(s)}_\mathrm{ur}}{1-\rho_{ss}^{(s)2}}\right)^{2m+2n+x}}\notag \\&\!\!\!\times\!\!{}_1F_1^2\!\!\left(\!\!n\!+\!m\!+\!\frac{x}{2},\!n\!-\!1\!+\!x, \!\frac{\zeta^2\!\kappa_\mathrm{ur}^{(s)}\!\!\left(\!1\!+\!|\rho_{ss}^{(s)}|^2\!-\!2\mu_c\!\right)\!\!\!\left(\!1\!-\!\rho_{ss}^{(s)2\!}\right)}{4\left(1\!+\!\kappa_\mathrm{ur}^{(s)}\right)}\!\!\right)\!\!,
\end{align}
and $\mathbf{C}(x)$ is given in (\ref{eq:Cx}), $\rho^{(k)}_{ss}\!=\! \mathbf{R}^{(k)}_{\mathrm{ur},s,s,i,j}$, $r_{ss}^{(k)}=\mathbf{R}^{(k)}_{\mathrm{r},s,s,i,j}$, $\rho^{(k)}_{st}=\mathbf{R}^{(k)}_{\mathrm{ur},s,t,i,j}$, $r^{(k)}_{st}=\mathbf{R}^{(k)}_{\mathrm{r},s,t,i,j}$, $\rho^{(s)}_{ss}=\mathbf{R}^{(s)}_{\mathrm{ur},s,s,i,j}$, $\mu_c\!=\!\rho^{(s)}_{ss}\!\cos\!\left(\!\angle\mathbf{a}_{\mathrm{ur},s,j}^{(s)}\!-\!\angle\mathbf{a}_{\mathrm{ur},s,i}^{(s)}\!\right)$, $\mu_s\!=\!\rho^{(s)}_{ss}\!\sin\!\left(\!\angle\mathbf{a}_{\mathrm{ur},s,j}^{(s)}\!-\!\angle\mathbf{a}_{\mathrm{ur},s,i}^{(s)}\!\right)$, $\kappa_0\!=\!\frac{2|\rho^{(s)}_{ss}|}{1\!-\!|\rho^{(s)}_{ss}|^2}\!\!\left(\!1\!+\!\kappa_\mathrm{ur}^{(s)}\!\right)$, $\zeta=\frac{2\sqrt{1+\kappa_\mathrm{ur}^{(s)}}}{1-|\rho^{(s)}_{ss}|^2}$, $\phi\!=\!\angle(1\!-\!\mu_c\!+\!j\mu_s)\!+\!\!\angle\mathbf{a}_{\mathrm{ur},s,j}^{(s)}\!\!-\!\!\angle(1\!-\!\mu_c\!-\!j\mu_s)\!-\!\!\angle\mathbf{a}_{\mathrm{ur},s,i}^{(s)}\!+\!\angle\rho^{(s)}_{ss}\!$, $\epsilon_0 = 1$, and $\epsilon_n = 2$ when $n\geq 1$ from \cite{mendes_general_2007}. 
\begin{remark}
    $\mathbb{E}[\mathbf{g}_\mathrm{k}^\dagger\mathbf{g}_\mathrm{k}]$ represents the sum of the gains of the channels through subsurfaces designed for other \glspl{UE}. Increased correlation between elements of any subsurface is still beneficial to \gls{UE} $k$, again due to the presence of Gaussian hypergeometric functions that increase monotonically with the spatial correlation of \gls{RIS} elements. Therefore, it is beneficial to all \glspl{UE} if all \gls{RIS} elements are closely collocated.
\end{remark}
\begin{remark}
     As the terms in (\ref{eq:Efkfk}) and (\ref{eq:Egkgk}) are quadratic, and thus typically larger than the cross product terms, a change in spatial correlation at the \gls{RIS} is more influential to the overall \gls{SNR} than a change in spatial correlation at the \gls{BS}.
\end{remark}

Combining (\ref{eq:Ehdhd}), (\ref{eq:Ehdf}), (\ref{eq:Ehdgk}), (\ref{eq:Efkgk}), (\ref{eq:Efkfk}) and (\ref{eq:Egkgk}) with (\ref{eq:SNRterms}) gives the complete mean \gls{SNR}.

\subsection{Special Case 1: Mean SNR for Correlated Ricean $\mathbf{H}_\mathrm{rb}$ with Correlated Rayleigh UE Channels}
Now consider the $\mathbb{E}[\mathrm{SNR}_k]$ subcase where $\mathbf{H}_\mathrm{rb}$ is correlated Ricean and the \gls{UE} channels are correlated Rayleigh.

Assuming $\mathbf{h}^{(k)}_{\mathrm{d}} = \sqrt{\beta_\mathrm{d}^{(k)}}\Tilde{\mathbf{h}}_\mathrm{d}^{(k,\mathrm{SC})}$ and $\mathbf{h}^{(k)}_{\mathrm{ur}} = \sqrt{\beta_\mathrm{ur}^{(k)}}\Tilde{\mathbf{h}}_\mathrm{ur}^{(k,\mathrm{SC})}$ are correlated Rayleigh channels, then $\mathbb{E}[\mathbf{h}_{\mathrm{d}}^{(k)}] = \mathbb{E}[\mathbf{h}_{\mathrm{ur}}^{(k)}] = \mathbf{0}$. As $\mathbf{h}^{(k)}_{\mathrm{d}}$ and $\mathbf{h}^{(k)}_{\mathrm{ur}}$ are independent, and $\mathbf{h}_{\mathrm{ur}}^{(k)}$ and $\mathbf{h}_{\mathrm{ur}}^{(s)}$ are independent for $s \neq k$ , $\mathbb{E}[\mathbf{f}_k^\dagger\mathbf{g}_k]$ and $\mathbb{E}[\mathbf{h}^{(k)\dagger}_{\mathrm{d}}\mathbf{g}_\mathrm{k}]$ contain zero-mean terms. Therefore, (\ref{eq:SNRterms}) simplifies to
\begin{equation}
    \mathbb{E}[\mathrm{SNR}_k]\!=\!\frac{E_s}{\sigma^2}\mathbb{E} \Big[ \mathbf{h}^{(k)\dagger}_{\mathrm{d}} \mathbf{h}^{(k)}_{\mathrm{d}}\!+2\Re(\mathbf{h}_{\mathrm{d}}^{(k)\dagger} \mathbf{f}_k)\!+\!\mathbf{f}_k^\dagger \mathbf{f}_k\!+\! \mathbf{g}_k^\dagger \mathbf{g}_k\Big]\!. \label{eq:ESNR_simplified}
\end{equation}
The result for $\mathbb{E}[\mathbf{h}^{(k)\dagger}_{\mathrm{d}} \mathbf{h}^{(k)}_{\mathrm{d}}]$ is the same as (\ref{eq:Ehdhd}), while (\ref{eq:Ehdf}), (\ref{eq:Efkfk}) and (\ref{eq:Egkgk}) no longer rely on UE link K-factors and can be simplified to
\begin{equation}
    \mathbb{E}[\mathbf{h}^{(k)\dagger}_{\mathrm{d}}\mathbf{f}_\mathrm{k}] = \frac{N_k\pi||\mathbf{R}^{(k)1/2}_{\mathrm{d}}\mathbf{a}^{(k)}_\mathrm{b}||\eta_\mathrm{rb}^{(k)}\sqrt{\!\beta^{(k)}_{\mathrm{d}}\!\beta^{(k)}_{\mathrm{rb}}\!\beta^{(k)}_{\mathrm{ur}}}}{4}, \label{eq:Ehdf_simplified}
\end{equation}
\begin{multline}
    \mathbb{E}[\mathbf{f}_\mathrm{k}^\dagger\mathbf{f}_\mathrm{k}] = M\beta_\mathrm{rb}^{(k)}\beta_\mathrm{ur}^{(k)}\bigg[N_k\left(\eta_\mathrm{rb}^{(k)2}+\zeta_\mathrm{rb}^{(k)2}\right) + \frac{\pi}{4}\sum_{i=1}^{N_k}\sum_{j\neq i} \\ \times\!\left(\!\eta_\mathrm{rb}^{(k)2}\!\!+\!\zeta_\mathrm{rb}^{(k)2}\!\mathbf{A}_{i,j}\!\right)\!{}_2F_{1}\!\left(\frac{-1}{2},\frac{-1}{2},1,|\rho_{kk}^{(k)}|^2\!\right)\!\!\bigg], \label{eq:Efkfk_simplified}
\end{multline}
\begin{multline}
    \mathbb{E}[\mathbf{g}_\mathrm{k}^\dagger\mathbf{g}_\mathrm{k}] =M\beta_\mathrm{rb}^{(k)}\beta_\mathrm{ur}^{(k)}\frac{\pi}{4}\sum_{s\neq k}\sum_{i=1}^{N_k}\sum_{j=1}^{N_k}\Big(\eta_\mathrm{rb}^{(k)2}\mathbf{a}_{\mathrm{r},s,i}^{(k)}\mathbf{a}_{\mathrm{r},s,j}^{(k)*} \\ + \zeta_\mathrm{rb}^{(k)2}r_{ss}^{(k)}\!\Big)\rho_{ss}^{(k)}\rho_{ss}^{(s)}\mathbf{a}_{\mathrm{r},s,i}^{(s)*}\mathbf{a}_{\mathrm{r},s,j}^{(s)}{}_2F_1\left(\frac{1}{2},\frac{1}{2},2,|\rho_{ss}^{(s)}|^2\right)\!. \label{eq:Egkgk_simplified}
\end{multline}

Substituting (\ref{eq:Ehdhd}), (\ref{eq:Ehdf_simplified}), (\ref{eq:Efkfk_simplified}) and (\ref{eq:Egkgk_simplified}) into (\ref{eq:ESNR_simplified}) gives the mean \gls{SNR}.
\begin{remark}
\label{rem:riceray}
    The contribution in \eqref{eq:Ehdf_simplified} is enhanced by alignment between the scattering in the direct link and \gls{LoS} component of the \gls{RIS} link. Specifically, alignment between the leading eigenvectors of $\mathbf{R}_\mathrm{d}^{(k)}$ and $\mathbf{a}_\mathrm{b}^{(k)}$ increases the \gls{SNR}. In \eqref{eq:Efkfk_simplified}, increasing channel power enhances the dominant first term. The second summation term tends to increase with spatial correlation due to the monotonically increasing Gaussian hypergeometric function, while the summation over $\mathbf{A}_{i,j}$ terms is less significant due to the cancellation inherent in summing over phases. The same pattern occurs in \eqref{eq:Egkgk_simplified} with increasing correlation boosting the positive $i=j$ component and $i\neq j$ terms tending to cancel out.
\end{remark}

\subsection{Special Case 2: Mean SNR for LoS $\mathbf{H}_\mathrm{rb}$ with Correlated Rayleigh UE Channels}
Now consider the further simplification of $\mathbb{E}[\mathrm{SNR}_k]$ for \gls{LoS} $\mathbf{H}_\mathrm{rb}$ and correlated Rayleigh \gls{UE} links.

Again assuming $\mathbf{h}^{(k)}_{\mathrm{d}} = \sqrt{\beta_\mathrm{d}^{(k)}}\Tilde{\mathbf{h}}_\mathrm{d}^{(k,\mathrm{SC})}$ and $\mathbf{h}^{(k)}_{\mathrm{ur}} = \sqrt{\beta_\mathrm{ur}^{(k)}}\Tilde{\mathbf{h}}_\mathrm{ur}^{(k,\mathrm{SC})}$, let $\mathbf{H}_\mathrm{rb}^{(k)}=\sqrt{\beta_\mathrm{rb}^{(k)}}\Tilde{\mathbf{H}}_\mathrm{rb}^{(k,\mathrm{LoS})}$. Due to the independence of $\mathbf{h}^{(k)}_{\mathrm{d}}$ and $\mathbf{h}^{(k)}_{\mathrm{ur}}$, the $\mathbb{E}[\mathrm{SNR}_k]$ expression is equivalent to (\ref{eq:ESNR_simplified}). $\mathbb{E}[\mathbf{h}^{(k)\dagger}_{\mathrm{d}} \mathbf{h}^{(k)}_{\mathrm{d}}]$ is again equivalent to (\ref{eq:Ehdhd}), while the remaining terms are no longer dependent on the \gls{RIS}-\gls{BS} link K-factor and can be further simplified to
\begin{equation}
    \mathbb{E}[\mathbf{h}_{\mathrm{d}}^{(k)\dagger} \mathbf{f}_k] = \frac{N_k\pi||\mathbf{R}^{(k)1/2}_\mathrm{d}\mathbf{a}^{(k)}_\mathrm{b}|| \sqrt{\beta^{(k)}_{\mathrm{d}}\beta^{(k)}_{\mathrm{rb}}\beta^{(k)}_{\mathrm{ur}}}}{4}, \label{eq:Ehdf_simplified2}
\end{equation}
\begin{equation}
    \mathbb{E}[\mathbf{f}_k^\dagger \mathbf{f}_k]\!=\!M\!\beta_{\mathrm{rb}}^{(k)}\!\beta_{\mathrm{ur}}^{(k)}\!\bigg(\!\!N_k\!+\!\frac{\pi}{4}\!\sum\limits_{i=1}^{N_k}\!\sum\limits_{j\neq i}{}_2F_1\!\!\left(\!\!\frac{-1}{2},\frac{-1}{2};1;|\rho_{kk}^{(k)}|^2\!\right)\!\!\!\bigg), \label{eq:Efkfk_simplified2}
\end{equation}
\begin{multline}
    \mathbb{E}[\mathbf{g}_\mathrm{k}^\dagger\mathbf{g}_\mathrm{k}] =M\beta_\mathrm{rb}^{(k)}\beta_\mathrm{ur}^{(k)}\frac{\pi}{4}\sum_{s\neq k}\sum_{i=1}^{N_k}\sum_{j=1}^{N_k}\rho_{ss}^{(k)}\rho_{ss}^{(s)}\mathbf{a}_{\mathrm{r},s,i}^{(k)}\mathbf{a}_{\mathrm{r},s,j}^{(k)*}  \\ \times\mathbf{a}_{\mathrm{r},s,i}^{(s)*}\mathbf{a}_{\mathrm{r},s,j}^{(s)}{}_2F_1\left(\frac{1}{2},\frac{1}{2},2,|\rho_{ss}^{(s)}|^2\right)\!. \label{eq:Egkgk_simplified2}
\end{multline}

Substituting (\ref{eq:Ehdhd}), (\ref{eq:Ehdf_simplified2}), (\ref{eq:Efkfk_simplified2}) and (\ref{eq:Egkgk_simplified2}) into (\ref{eq:ESNR_simplified}) gives the mean \gls{SNR}, matching those results we first reported in \cite{inwood_phase_2023}.

\begin{remark}
     As in Remark \ref{rem:riceray}, alignment between $\mathbf{R}_\mathrm{d}^{(k)}$ and $\mathbf{a}_\mathrm{b}^{(k)}$ and increased spatial correlation tends to increase the \gls{SNR} components in \eqref{eq:Ehdf_simplified2}-\eqref{eq:Egkgk_simplified2}.
\end{remark}

\section{Numerical Results}
\label{sec:numres}
Numerical results were generated to verify the analysis above and explore the performance of the subsurface design. The channel gain values are selected based on the distance based path loss model detailed in \cite{wu_intelligent_2019}, where
\begin{equation}
    \beta =  C_0\left({d}/{D_0}\right)^{-\alpha},
\end{equation}
$D_0$ is the reference distance of 1 m, $C_0$ is the path loss at $D_0$ (-30 dB), $d$ is the link distance in metres and $\alpha$ is the path loss exponent ($\alpha_{\mathrm{d}} = \alpha_{\mathrm{rb, NLoS}} =3.5$, $\alpha_{\mathrm{rb,LoS}} = 2$ and $\alpha_{\mathrm{ru}} = 2.8$). 

\begin{figure}[ht]
    \includegraphics[trim={1.25cm 2.3cm 0.1cm 0.04cm},clip,scale=0.37]{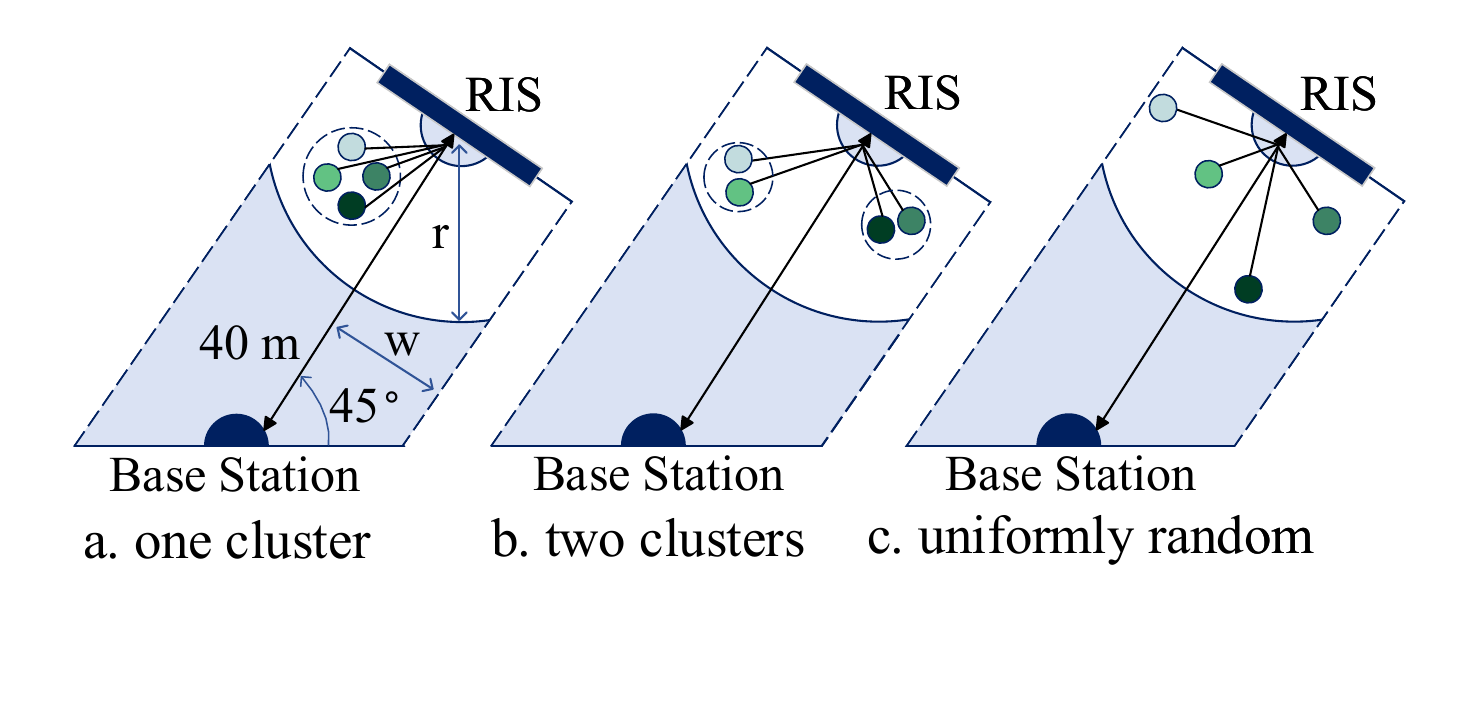}
    \caption{User drop configurations for one cluster (a), two clusters (b) and random (c).}
    \label{fig:system_layout}
\end{figure}

It is assumed that the \gls{RIS} is fixed at 40 m away from the \gls{BS}, at an angle of $\frac{\pi}{4}$ rad. Users are then dropped in a straight corridor of width $2w$ between \gls{BS} and \gls{RIS}, at a maximum distance of $r$ from the \gls{RIS} with an exclusion zone of 1 m around the \gls{RIS}. This region ensures that the path from the \gls{UE} through the \gls{RIS} significantly contributes to the total channel, and the effects of the \gls{RIS} can be clearly seen. User clusters are considered to investigate the effect of people naturally congregating. Three types of user grouping were used in simulations. As seen in Fig. \ref{fig:system_layout}, layout A places all $K$ \glspl{UE} in one cluster, where they are all in the same area spaced 1 m apart and layout B creates 2 clusters of $\frac{K}{2}$ \glspl{UE}, with users spaced 1 m apart. Cluster locations are drawn from a uniform distribution. Layout C places each of the $K$ \glspl{UE} randomly within the corridor, again using a uniform distribution. To ensure that results are general and not tied to a specific \gls{UE} location, $10^4$ random user drops are generated per layout, each with $10^6$ replicates.

For simulation purposes, we use the Rayleigh fading correlation model proposed in \cite{bjornson_rayleigh_2021}, where
\begin{equation}
    \mathbf{R}_{n,m} = \mathrm{sinc} \left( 2 \,d_{mn} \right),\quad n,m = 1, \dots, L,
\end{equation}
$\mathbf{R}\in\{\mathbf{R}_\mathrm{d},\mathbf{R}_\mathrm{b},\mathbf{R}_\mathrm{r},\mathbf{R}_\mathrm{ur}\}$, $d_{mn}$ is the Euclidean distance between \gls{BS} antennas/ \gls{RIS} elements $m$ and $n$, measured in wavelength units, and $L$ is the number of \gls{BS} antennas/ \gls{RIS} elements. 

The steering vectors $\mathbf{a}_\mathrm{d}$, $\mathbf{a}_\mathrm{b}$, $\mathbf{a}_\mathrm{r}$ and $\mathbf{a}_\mathrm{ur}$ correspond to the \gls{VURA} model and are given in  \cite{miller_analytical_2019}. Elements are arranged in a \gls{VURA} at intervals of $d_b$ and $d_r$ wavelengths at the \gls{BS} and \gls{RIS}, respectively. $M_x$ and $N_x$ are the number of \gls{BS} and \gls{RIS} elements per row and $M_z$ and $N_z$ are the number of \gls{BS} and \gls{RIS} elements per column, such that $M = M_xM_z$ and $N = N_xN_z$. $\theta_\mathrm{D}$ and $\phi_\mathrm{D}$ are the elevation and azimuth \glspl{AoD} at the \gls{RIS} and $\theta_\mathrm{A}$ and $\phi_\mathrm{A}$ are the corresponding elevation and azimuth \glspl{AoA} at the \gls{BS}. We assume the \gls{RIS} is on a $\frac{\pi}{4}$ rad angle with respect to the \gls{BS}, so $\phi_\mathrm{D} = \frac{5\pi}{4}$ rad and $\phi_\mathrm{A} = \frac{\pi}{4}$ rad. We also assume both are at the same height, so $\theta_\mathrm{D}=\theta_\mathrm{A}=\frac{\pi}{2}$ rad. 

Without loss of generality we assume $\sigma^2=1$ and $E_s$ is selected in Sec \ref{sec:numres_sub} so that the mean output \gls{SNR} for one \gls{UE} is 5 dB in a system using the \gls{SD} where $N=128$, $M=16$, $d_r = 0.1$, $d_b=0.5$ and $\kappa_{d} = \kappa_{rb} =\kappa_{ur} =1$. In Sec. \ref{sec:numres_sub2}, we select $E_s$ so that the mean output \gls{SNR} for one \gls{UE} is 5 dB in a system using the \gls{SD} where $N=128$, $M=16$, $d_r = 0.1$, $d_b=0.5$, $\kappa_\mathrm{ur}=\kappa_\mathrm{d}=1$ and the \gls{RIS}-\gls{BS} channel is purely \gls{LoS}, an important special case. These parameter values and definitions do not change throughout the results, unless specified otherwise.

\subsection{Mean SNR Comparison of Subsurface Methods in Sec. \ref{sec:phase_selection_methods}}
\label{sec:numres_sub}

\begin{figure}[ht]
    \centering
    \includegraphics[scale=0.57]{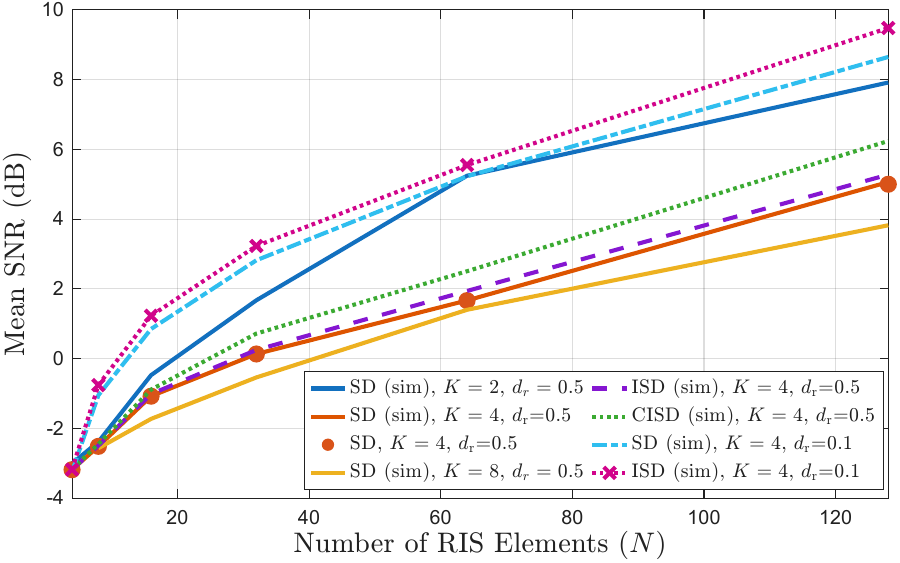}
    \caption{Comparison of mean \gls{SNR} for a randomly selected \gls{UE} for different subsurface phase selection methods while varying the number of \gls{RIS} elements.}
    \label{fig:ESNR_vs_actual}
\end{figure}

Fig. \ref{fig:ESNR_vs_actual} verifies the analytical results in Sec. \ref{sec:ESNR_Ricean} by showing the equivalence of $\mathbb{E}[\mathrm{SNR}]$ and the simulated mean \gls{SNR} for the \gls{SD}, and then compares this with the simulated mean \gls{SNR} for the \gls{ISD}. Two versions of the \gls{ISD} were implemented. The first iterates through all subsurfaces once, as shown in Sec. \ref{sec:ISD}. The second iterates through all subsurfaces and calculates the \gls{SNR} of the system. It then repeats this process multiple times until the \gls{SNR} of the system has converged, i.e. it increases by less than a specified tolerance from one iteration to the next. For these simulations, the tolerance was specified to be $10^{-4}$. This method is referred to as the \gls{CISD}. The array parameters for Fig. \ref{fig:ESNR_vs_actual} are $M = 16$, $M_x = 4$, $d_b = 0.5$ and all K-factors are 1.

It can be seen that the simulated mean \gls{SNR} for the method outlined in Sec. \ref{sec:LOS_phase_selection} is virtually indistinguishable from the $\mathbb{E}[\mathrm{SNR}]$ derived in Sec. \ref{sec:ESNR_Ricean}, verifying that result. As expected, the mean \gls{SNR} for all 3 methods increases with $N$, however the rate of improvement decreases as $N$ increases. This is expected, as it is well known that the \gls{RIS} \gls{SNR} can grow quadratically with $N$ \cite{singh_optimal_2021} so that the dB relationship is logarithmic. The quadratic behavior can be seen in (\ref{eq:Efkfk}), (\ref{eq:Efkfk_simplified}), and (\ref{eq:Efkfk_simplified2}) where the $\mathbb{E}[\mathbf{f}_k^\dagger\mathbf{f}_k]$ term contains the sum of $N_k(N_k-1)$ positive terms, giving a contribution of $\mathcal{O}(N_k^2)$. The \gls{ISD} and converged \gls{ISD} both provide more benefit at higher $N$. At high $N$, more improvement can be made by iterating, as there are more elements to vary and hence more opportunities to improve the performance of the system. This improvement is more pronounced when the correlation between \gls{RIS} elements is lower. Figure \ref{fig:ESNR_vs_actual} shows that the gap between the mean \gls{SNR} of the \gls{SD} and \gls{CISD} is larger for \gls{RIS} element spacings of $d_r = 0.5$ than for $d_r=0.1$. This may be due to the fact that there is more variability in the previously set subsurfaces creating the fixed channels for the iterative design. Hence, reduced correlation leads to greater variation in the fixed channels, which the iterative design can exploit. Additionally, when $N$ is kept constant, the average \gls{SNR} per \gls{UE} decreases as the number of \glspl{UE} increases. This reduction arises because the fixed set of \gls{RIS} elements must be divided between more \glspl{UE}, resulting in fewer elements being allocated to each \gls{UE}.

A key design choice for the \gls{ISD} is the order in which subsurfaces are set. The standard \gls{ISD} in Sec. \ref{sec:ISD} sets the subsurface of weaker \glspl{UE} first. This allows the strongest \glspl{UE} to benefit from additional channel information, and leads to higher average performance. However, this does not lead to fair performance across all UEs. Table \ref{tab:isd_v_risd} compares the \gls{ISD} to two other versions. The first is a reverse version, where strongest \glspl{UE} are set first, allowing the weakest \glspl{UE} to benefit from extra information. The second is a random version, where the order in which the \glspl{UE} are set is random. This version acts as a benchmark and allows insights into the respective benefits provided by the \gls{ISD} and reverse \gls{ISD}. The mean \gls{SNR} of the strongest \gls{UE} (\gls{UE} 1) to the weakest \gls{UE} (\gls{UE} 4) for all three methods is shown for $N=128$ and $d_r=0.5$.
\begin{table}[ht]
    \centering
    \caption{Mean \gls{SNR} ($\mathrm{dB}$) for each \gls{UE} using the \gls{ISD} and the reverse \gls{ISD}.}
    \begin{tabular}{|c|c c c c|}
        \hline & UE 1 & UE 2 & UE 3 & UE 4 \\
        \hline ISD & 15.12 & 7.87 & 3.35 & 0.30\\
        Reverse ISD & 13.38 & 7.25 & 3.44 & 0.93\\ 
        Random ISD & 14.42 & 7.51 & 3.35 & 0.63 \\ \hline
        
    \end{tabular}
    \label{tab:isd_v_risd}
\end{table}

The reverse \gls{ISD} offers fairer performance at the expense of a rate drop for the stronger \glspl{UE} and a lower average rate. The random \gls{ISD} provides a balance between the two, with less of a rate drop than the reverse \gls{ISD}, but also lower fairness. Thus, the choice of subsurface ordering should depend on whether fairness or average rate is the priority.

The increase in performance provided by the \gls{ISD} and \gls{CISD} also leads to an increase in required \gls{CSI}. Both methods require the same amount of \gls{CSI} as each needs to estimate $N$ channels between the \gls{RIS} and the \gls{UE}, as opposed to the $\frac{N}{K}$ required by the \gls{SD}. This is due to the need to measure the impact of each subsurface on all users, rather than on just the user it is designed for. Gathering \gls{CSI} is expected to be challenging for \gls{RIS} systems, so the benefit from an iterative method would need to be weighed up against this additional \gls{CSI} requirement. 

Each iteration of the \gls{ISD} is very simple. Additional complex multiplies are needed to compute channel norms to order subsurfaces by impact and to recalculate fixed channel components. For the \gls{CISD}, the \gls{SNR} also needs to be computed each iteration. Table \ref{tab:SDcomplexity} compares the complexity of the three subsurface methods if $n$ iterations are required. 
\begin{table}[ht]
    \centering
    \caption{Complexity comparison of the subsurface methods.}
    \setlength\tabcolsep{1.5pt} 
    \begin{tabular}{|C{16mm}|c|C{25mm}|C{25mm}|}
         \hline
         & \textbf{SD} & \textbf{ISD} & \textbf{CISD} \\
         \hline
        Multiplies & $KM\!+\!N$ & $NK\!+\! n\!\Big(\!\frac{(K\!-\!1)!MN}{K}$ $\times\!\!\left(\!\frac{N}{K}\!+\!1\!\right)\!+\!KM\!+\!N\!\Big)$ & $NK\!+\! n\!\Big(\!\frac{(K\!-\!1)!MN}{K}$ $\times\!\!\left(\!\frac{N}{K}\!+\!1\!\right)\!+\!KM\!(2\!+\!N(N\!+\!1)\!)\!+\!N\Big)$\\
        \hline
        Divides & $2N+K$ & $n(2N+K)$  & $n(2N+K)$ \\
        \hline
        Mathematical Functions  & None & 1 sort($\cdot$) \newline 1 index($\cdot$)& 1 sort($\cdot$) \newline 1 index($\cdot$)\\
        \hline
        Array \newline Functions  & None & 1 sort($\cdot$) \newline 1 index($\cdot$)& 1 sort($\cdot$) \newline 1 index($\cdot$)\\
        \hline
    \end{tabular}
    \label{tab:SDcomplexity}
\end{table}

\begin{figure}[ht]
    \centering
    \includegraphics[scale=0.57]{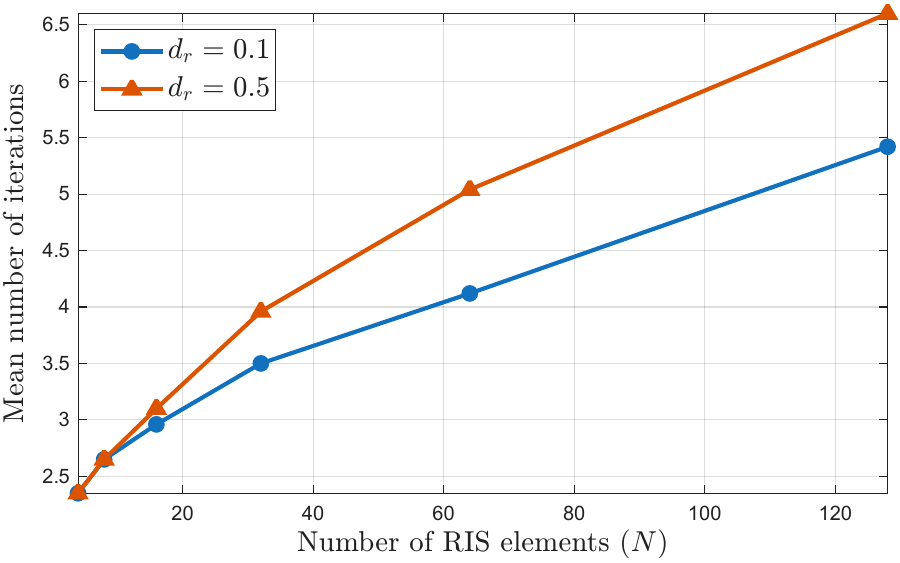}
    \caption{Mean number of iterations of the \gls{ISD} required for convergence.}
    \label{fig:num_its}
\end{figure}

The average number of iterations required for convergence for each $N$ is shown in Fig. \ref{fig:num_its}. The largest mean number of iterations for an $M=16$, $N = 128$ system occurs when $d_r=0.5$ and is 6.60. In this scenario, when the maximum number of iterations is fixed at 7, the resulting \gls{SNR} averages to be 99.8\% of the fully converged \gls{SNR}. Hence, restricting the number of iterations to a fixed amount has little impact on performance.

\subsection{Mean Sum Rate of SD versus Non-Subsurface Methods}
\label{sec:numres_sub2}
Many existing \gls{MU} \gls{RIS} phase selection methods, such as those in \cite{abeywickrama_intelligent_2020, gao_robust_2021, buzzi_ris_2021, chen_low_2024}, focus on achieving near optimal performance. Necessarily, they have challenging computational complexity requirements. The lowest complexity \gls{MU} method we are aware of is that proposed in \cite{singh_efficient_2022}. While other methods may offer better performance, the design in \cite{singh_efficient_2022} is more likely to be considered for practical implementation. This method approximately minimizes the \gls{TMSE} of a \gls{MMSE} receiver. All users are placed in one band, so each user has $K$ times the \gls{BW} when compared to users served by the SD. This discrepancy in available \gls{BW} per \gls{UE} would exist between the \gls{SD} and any \gls{MU} approach where all of the \glspl{UE} are in a single frequency band. We believe it is important to compare our method to this type of scenario, as this is the most common type of \gls{MU} \gls{RIS} design in the literature. Also, considering a key benefit of the \gls{SD} is a significant complexity reduction, a low-complexity \gls{MU} approach was considered to be a fair comparison. To show the low complexity nature of the \gls{TMSE} method and the significant complexity reduction afforded by the \gls{SD}, the number and complexity of matrix operations required for both methods is compared to the low-complexity near-optimal approach in \cite{chen_low_2024} in Table \ref{tab:compcomp}. The approach in \cite{chen_low_2024} jointly optimizes the \gls{RIS} phase shifts, precoder and combiner for a single \gls{UE} system to minimize the \gls{MSE}. For a fair comparison with the \gls{SD} and \gls{TMSE} approaches, the number of operations for the single \gls{UE} case are scaled by $K$, representing an \gls{OFDM}-type system with no \gls{IUI}.
\begin{table}[t]
    \setlength{\tabcolsep}{3pt}
    \centering
    \caption{Complexity comparison of the \gls{SD}, the \gls{TMSE} method in \cite{singh_efficient_2022} and the \gls{MSE} method in \cite{chen_low_2024}.}
    \begin{tabular}{|C{1.6cm}|C{1.6cm}|C{2cm}|C{1.6cm}|}
        \hline & \textbf{SD} & \textbf{TMSE} \cite{singh_efficient_2022} & \textbf{MSE} \cite{chen_low_2024} \\ \hline
        Complex Multiplicative Operations & $3N\!+\! K(M\!+\!1)$ & $(N\!+\!M)(K\!+\!K^2)\!+\!K(3\!+\!3K+\!2K^2)$ & $ nK\!(2N(M\!+\!1)\!+\!M \!+\!3)$ \\ \hline
        Matrix Functions & None & 2 $K\!\times\! K$ inverses \newline
        1 $K\!\times\! K$ eigendecomposition & 1 $M\times1$ eigendecomposition\\ \hline
    \end{tabular}
    \label{tab:compcomp}
\end{table}

Assuming that $N$ is the largest variable, the \gls{SD} offers a multiplicative reduction of a factor of $K+K^2$ over the \gls{TMSE} method, and of $nKM$ over the \gls{MSE} approach. For a system where $N=128$, $M=32$, $K=4$ and $n=40$ (as suggested in \cite{chen_low_2024}), the \gls{SD}, \gls{TMSE} method and \gls{MSE} method require 516, 3388 and 1356960 multiplicative operations, respectively, highlighting the significant computational savings provided by the \gls{SD}. Additionally, the \gls{SD} does not require matrix functions, unlike the others. Results in \cite{inwood_phase_2023} also showed that the \gls{SD} provides fairer results according to Jain's fairness test, uses a much simpler receiver type (\gls{MF} vs \gls{MMSE}) and reduces \gls{CSI} requirements by a factor of $K$ when compared to the \gls{TMSE} approach. Hence, the \gls{SD} approach offers substantial complexity savings even compared to the simplest \gls{RIS} designs.

In this section, the mean sum rates for the \gls{SD} and \gls{TMSE} methods are compared with the mean sum rate for randomly selected phases. The inclusion of random phases creates a lower bound to compare both methods against. For the random approach, the phase shift of each element is generated from a uniform distribution between 0 and $2\pi$. It is assumed that all users are in one frequency band like the \gls{TMSE} example. We investigate the impact of important system features, such as K-factor levels, correlation, system size and \gls{UE} locations. \\

\subsubsection{K-Factors and Clustering}
\begin{figure}[ht]
    \centering
    \includegraphics[scale=0.57]{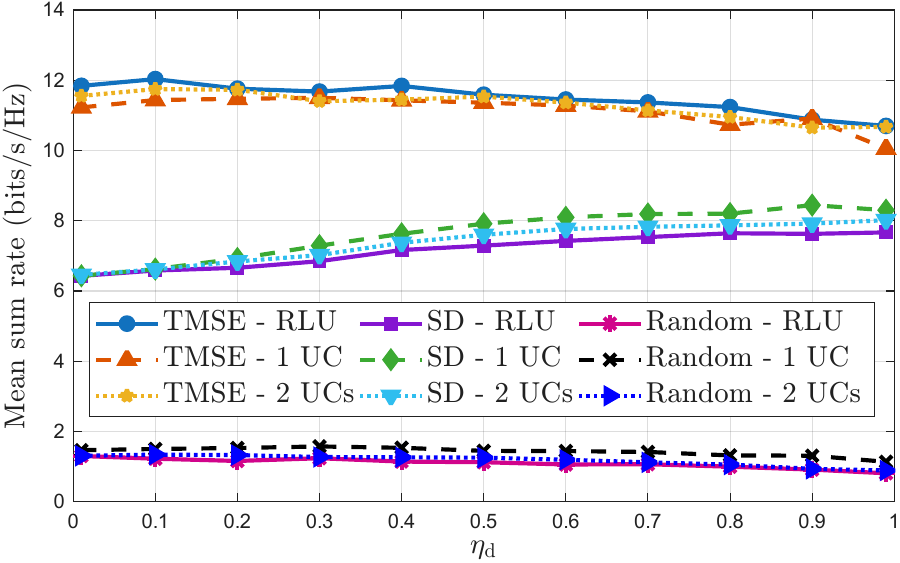}
    \caption{Comparison of mean sum rate for different phase selection methods while varying the \gls{UE}-\gls{BS} channel K-factor, and hence $\eta_\mathrm{d}$.}
    \label{fig:kfactor_d}
\end{figure}

\begin{figure}[ht]
    \centering
    \includegraphics[scale=0.57]{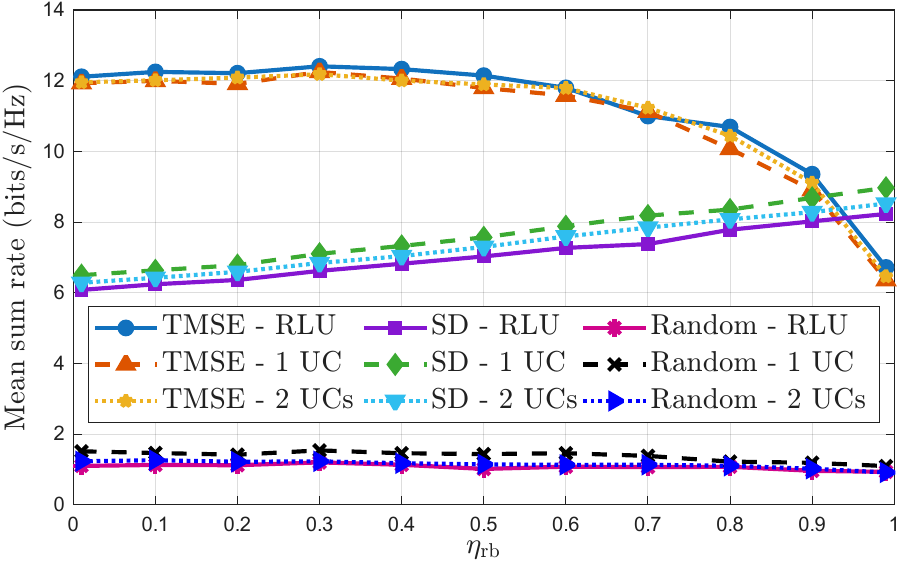}
    \caption{Comparison of mean sum rate for different phase selection methods while varying the \gls{RIS}-\gls{BS} channel K-factor, and hence $\eta_\mathrm{rb}$.}
    \label{fig:kfactor_rb}
\end{figure}

\begin{figure}[ht]
    \centering
    \includegraphics[scale=0.57]{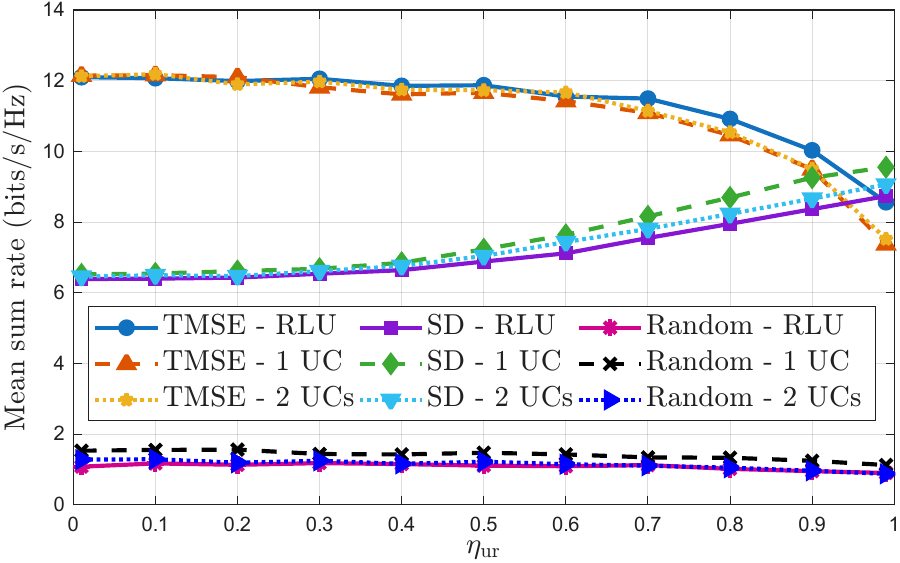}
    \caption{Comparison of mean sum rate for different phase selection methods while varying the \gls{UE}-\gls{RIS} channel K-factor, and hence $\eta_\mathrm{ur}$.}
    \label{fig:kfactor_ur}
\end{figure}

Figs. \ref{fig:kfactor_d}, \ref{fig:kfactor_rb} and \ref{fig:kfactor_ur} compare the mean sum rate for the \gls{SD} with that of random phases and the \gls{TMSE} design for varying levels of the \gls{UE}-\gls{BS}, \gls{RIS}-\gls{BS} and \gls{UE}-\gls{RIS} K-factor, respectively. $\eta$, as defined in (\ref{eq:etad}), (\ref{eq:etarb}) and (\ref{eq:etaur}), rather than K-factor itself is plotted so the proportion of the link that is \gls{LoS} can be easily observed.

The array parameters for Figs. \ref{fig:kfactor_d}, \ref{fig:kfactor_rb} and \ref{fig:kfactor_ur} are $M\!=\!16$, $N\!=\!128$, $K\!=\!4$, $M_x\!=\!4$, $N_x\!=\!16$, $d_b\!=\!0.5$, and $d_r\!=\!0.1$. The two fixed K-factors in each plot are set to 1 so that the \gls{LoS} and \gls{NLoS} component strengths are equal. Note that RLU represents randomly located \glspl{UE} and 1 UC and 2 UCs represent one and two clusters of \glspl{UE}, respectively.

Remarkably, Figs. \ref{fig:kfactor_d}, \ref{fig:kfactor_rb} and \ref{fig:kfactor_ur} show that despite the significant \gls{BW} restriction and lower complexity, the \gls{SD} actually outperforms the \gls{TMSE} when the K-factor of one of the RIS links is high. A \gls{LoS} or near-\gls{LoS} \gls{RIS} link is a key operating scenario that would be deliberately selected in practice. \gls{RIS} are often motivated as a method of overcoming propagation challenges in the direct path, and therefore, it is expected that their placement could be selected to ensure a strong \gls{LoS} path to the \gls{BS} and/or \glspl{UE}. In such configurations, a high K-factor is not only realistic but desirable. Thus, the \gls{SD} excels in a scenario of significant practical relevance and is well positioned for use in deployed RIS systems. While the \gls{TMSE} method yields higher sum rates at lower K-factors, its complexity scales much more rapidly with system size than that of the SD. Therefore, even for lower K-factor values, the SD has benefits for larger systems.

A higher K-factor is beneficial for the \gls{SD} but negatively impacts the \gls{TMSE} design. As the K-factor, and therefore $\eta$ (as defined in (\ref{eq:etad}), (\ref{eq:etarb}) and (\ref{eq:etaur})) of a channel increases, the rank of that channel decreases until it reaches rank-1 when $\eta = 1$. This limits the spatial multiplexing ability of the channel by reducing diversity. Insufficient diversity leads to the inability to separate \gls{MU} channels. As all users are in separate bands for the \gls{SD}, a lack of channel diversity is not an issue. The \gls{SNR} improves as channels can be aligned so that more power is directed towards the user as the rank decreases. However, for the \gls{TMSE} design where all users are located in one frequency band, the inability to separate channels as the K-factor increases leads to a significant drop in rate. A slight exception to this trend can be seen in Fig. \ref{fig:kfactor_rb} between $\eta_\mathrm{rb} = 0$ and $\eta_\mathrm{rb} = 0.3$. Note that the \gls{TMSE} method assumes a \gls{LoS} component in $\mathbf{H}_\mathrm{rb}$, so increasing the K-factor when $\eta_\mathrm{rb}$ is low helps slightly, as the channel becomes more similar to that which the framework was built upon. However, above this threshold, the reduction in diversity dominates and the overall rate reduces. The inability to separate channels also negatively impacts the random design, but this impact is lower as it is already operating at a low rate.

Varying the K-factors for the \gls{UE}-\gls{RIS} and \gls{RIS}-\gls{BS} channels results in similar mean sum rate behaviour. Data through the \gls{RIS} must travel through both channels, so changing the diversity of one places limits on the other, and thus on the whole \gls{RIS} link. This is more prominent when the \gls{LoS} link is dominant - the mean sum rate for the \gls{TMSE} method drops dramatically, due to the direct link providing the only diversity. The same can be observed for the random method, but to lesser extent. In comparison, when the \gls{UE}-\gls{BS} channel becomes strongly \gls{LoS}, the mean sum rate for the \gls{TMSE} design is less impacted. As discussed at the start of Sec. \ref{sec:numres}, \glspl{UE} are dropped in locations that ensure a strong \gls{RIS} path, to ensure the effects of the \gls{RIS} are visible. Therefore, the rate remains high in this case due to the strong \gls{RIS} link being unaffected. 

The impact of placing all four users in a cluster (layout A) is most noticeable in strongly \gls{LoS} channels, where it improves the rate of the \gls{SD} and lowers the rate of the \gls{TMSE} method. Clusters of densely located \glspl{UE} are expected in realistic scenarios of highly populated areas, such as stadiums or transit hubs. Clustering users reduces channel diversity and negatively impacts the rate of the single frequency \gls{TMSE} method. A key benefit of the \gls{SD} is its robustness to clustered users. As it does not rely on spatial multiplexing, clustered users improve its performance due to the similarity of the \gls{LoS} paths, allowing those subsurfaces designed for other users to provide better signal enhancement. Interestingly, clustering also helps the random method. This is likely due to the fact that when the random phases happen to be beneficial for one user, they benefits all \glspl{UE} in the cluster. 
\\

\subsubsection{Correlation}
\begin{figure}[ht]
    \centering
    \includegraphics[scale=0.57]{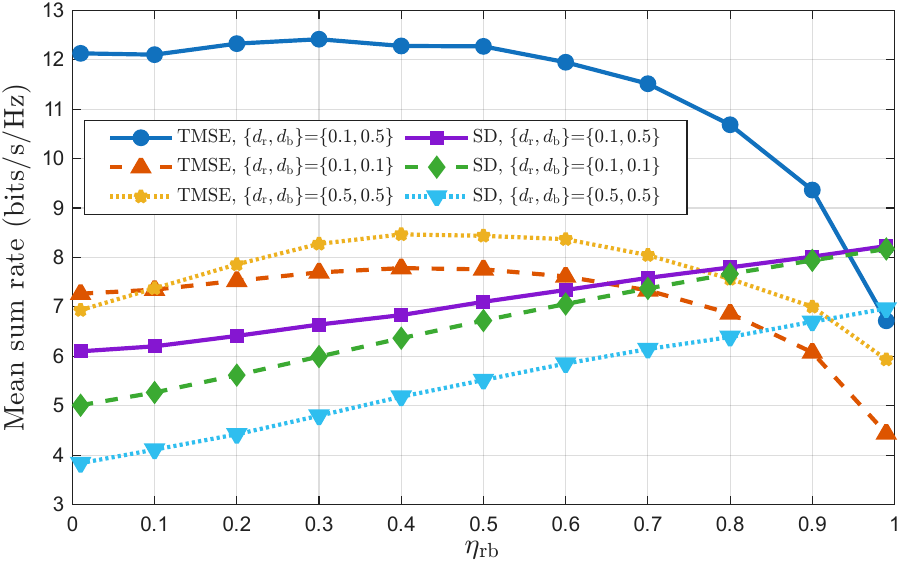}
    \caption{Comparison of mean sum rate for different correlations and phase selection methods while varying the \gls{RIS}-\gls{BS} channel K-factor, and hence $\eta_\mathrm{rb}$.}
    \label{fig:correlation_rb}
\end{figure}

Fig. \ref{fig:correlation_rb} compares the mean sum rate for the \gls{SD} and \gls{TMSE} designs for a range of element spacings at the \gls{RIS} and \gls{BS}. Elements located closer together result in higher correlation. As these results involve \gls{BS} element spacings of $d_b = 0.1\lambda$, the effects of mutual coupling at the \gls{BS} must be taken into account. Accordingly, for all curves corresponding to $d_b = 0.1\lambda$, the mutual coupling model in \cite{masouros_large_2013} is incorporated. Specifically, the spatial correlation matrices $\mathbf{R}_\mathrm{b}$ and $\mathbf{R}_\mathrm{d}$ are replaced by their effective counterparts $\mathbf{R}_\mathrm{b}^{\mathrm{eff}} = \mathbf{Z}^\dagger \mathbf{R}_\mathrm{b} \mathbf{Z}$ and $\mathbf{R}_\mathrm{d}^{\mathrm{eff}} = \mathbf{Z}^\dagger \mathbf{R}_\mathrm{d} \mathbf{Z}$, respectively, where $\mathbf{Z}$ is defined in \cite[Eq.~(9)]{masouros_large_2013}. The array parameters are the same as for Figs. \ref{fig:kfactor_d}-\ref{fig:kfactor_ur}, except for the varied element spacing.

For the \gls{SD}, it is observed that lower element spacing at the \gls{RIS} increases rate. This verifies the analysis in Sec. \ref{sec:ESNR_Ricean}, as (\ref{eq:Efkfk}) monotonically increases with correlation. In contrast, lower element spacing at the BS decreases rate. This also verifies findings in Sec. \ref{sec:ESNR_Ricean}, as (\ref{eq:Ehdf}) decreases with correlation. The \gls{RIS} element spacing is more influential than the \gls{BS} element spacing to the rate of the \gls{SD}. As expected, the rate is higher for the scenario where both spacings are $0.1\lambda$ than when both spacings are $0.5\lambda$, even when mutual coupling is considered at the \gls{BS}. This is due to \gls{RIS} correlation predominantly affecting the typically larger quadratic $\mathbf{f}_k^\dagger\mathbf{f}_k$ term in (\ref{eq:Efkfk}), and \gls{BS} correlation mostly affecting the typically smaller $\mathbf{h}_{\mathrm{d}}^\dagger\mathbf{f}_k$ cross product in (\ref{eq:Ehdf}).

The \gls{TMSE} method is designed for a \gls{RIS}-\gls{BS} channel with a strong \gls{LoS} component, which is assisted by higher correlation and a higher $\eta_\mathrm{rb}$. This is evidenced by smaller \gls{RIS} element spacing increasing the mean sum rate of the \gls{TMSE} design. However, as $\eta_\mathrm{rb}$ also becomes large, there is a substantial performance degradation. While the algorithm is more accurate for this scenario, strongly \gls{LoS} channels are not ideal to support multiple users, as the \gls{MU} channels cannot be separated. Also, unlike the \gls{SD}, the \gls{BS} element spacing is more influential to the rate of the \gls{TMSE} design. The \gls{TMSE} method relies on channel separability to serve multiple users in one band. Decreasing the \gls{BS} element spacing limits this in both the direct and \gls{RIS} paths, significantly decreasing the performance of the \gls{TMSE} method. The \gls{SD} performs comparatively better when correlation is high at the \gls{BS}, due to separability not being required for its operation.

\subsubsection{Array Dimensions}
\begin{figure}[ht]
    \centering
    \includegraphics[scale=0.57]{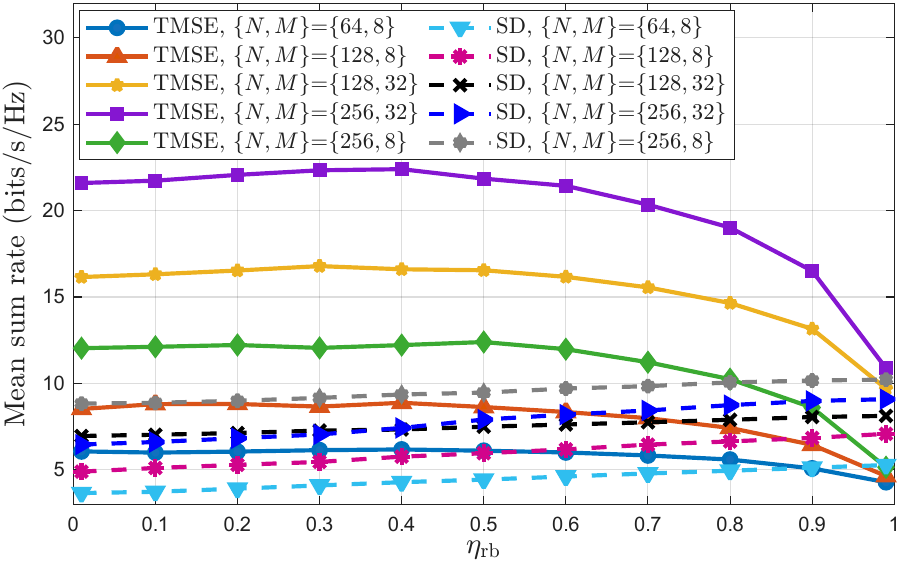}
    \caption{Comparison of mean sum rate for different array dimensions and phase selection methods while varying the \gls{RIS}-\gls{BS} channel K-factor, and hence $\eta_\mathrm{rb}$.}
    \label{fig:dims_rb}
\end{figure}

Fig. \ref{fig:dims_rb} compares the mean sum rate for 5 combinations of $M$ and $N$ values to investigate the impact of array dimensions. The array parameters are $K=4$, $d_b=0.5$, $d_r=0.1$, $\kappa_\mathrm{d} = 1$ and $\kappa_\mathrm{ur} = 1$.

In the case of \gls{TMSE}, it is observed that mean sum rate is influenced significantly more by $M$ than $N$. The largest mean sum rates are for the $M = 32$ cases, and then ordered by the size of $N$. Comparatively, the $M = 8$ cases give the smallest mean sum rates, with $M=8$, $N=64$ performing the worst. The \gls{TMSE} method needs sufficient $M$ to multiplex, as $M$ impacts both the direct and \gls{RIS} links. While a higher $N$ is beneficial, it only affects the \gls{RIS} link. Therefore, if $N$ is too low or $\eta_\mathrm{rb}$ is too high for efficient multiplexing in the \gls{RIS} path, the direct path could still support it if $M$ is large enough. Table \ref{tab:TMSE_pct} shows that the percentage decrease in the mean sum rate from $\eta_\mathrm{rb} = 0$ to $\eta_\mathrm{rb} = 1$ is greatest for the \gls{TMSE} method when $N$ is large and $M$ is small.

\begin{table}[ht]
    \centering
    \caption{Percentage decrease of \gls{TMSE} method mean sum rate from $\eta_\mathrm{rb} = 0$ to $\eta_\mathrm{rb} = 1$.}
    \label{tab:TMSE_pct}
    \begin{tabular}{|c|c|c|} 
    \hline
        \textbf{$M$} & \textbf{$N$} & \textbf{Mean sum rate decrease from $\eta_\mathrm{rb} = 0$ to $1$ (\%)} \\
         \hline
         8 & 256 & 56.2 \\
         32 & 256 & 49.6\\
         8 & 128 & 45.6\\
         32 & 128 & 40.0\\
         8 & 64 & 29.4\\
         \hline
    \end{tabular}
\end{table} 

In contrast, the mean sum rate of the \gls{SD} increases slightly with $\eta_\mathrm{rb}$ and is much more robust in low $M$ scenarios, outperforming the \gls{TMSE} method for all $M=8$ cases when $\eta_\mathrm{rb} \geq 0.9$. $N$ is the more influential parameter for the \gls{SD} - the highest mean sum rates are from the highest $N$ scenarios, and ordered within these by the highest $M$, which verifies the analysis in Sec. \ref{sec:numres}. As the \gls{SD} does not rely on multiplexing, $M$ is less important, and higher $N$ leads to more elements designed for each user and more scattering elements. A higher $\eta_\mathrm{rb}$ also leads to more power aligned in a specific direction and thus an increased mean sum rate. As percentage increase is a poor metric due to the small magnitude of the change, Table \ref{tab:SD_inc} shows the \gls{SD} mean sum rate increase from  $\eta_\mathrm{rb} = 0$ to $\eta_\mathrm{rb} = 1$ for each parameter set.

\begin{table}[ht]
    \centering
    \caption{Mean sum rate increase of \gls{SD} from $\eta_\mathrm{rb} = 0$ to $\eta_\mathrm{rb} = 1$.}
    \label{tab:SD_inc}
    \begin{tabular}{|c|c|c|} 
    \hline
        \textbf{$M$} & \textbf{$N$} & \textbf{Mean sum rate increase from $\eta_\mathrm{rb}\!=\!0$ to $1$ (bits/s/Hz)} \\
         \hline
         8 & 256 &  2.62\\
         8 & 128 & 2.18 \\
         8 & 64 & 1.62\\
         32 & 256 & 1.39\\
         32 & 128 & 1.16\\
         \hline
    \end{tabular}
\end{table}

The mean sum rate increase for the \gls{SD} is largest for small $M$ and large $N$. Changing $\eta_\mathrm{rb}$ has no effect on the direct link. Therefore, when the direct link is weaker and the \gls{RIS} link is stronger, a higher percentage of the total channel is being strengthened by the stronger \gls{LoS} component in the \gls{RIS}-\gls{BS} channel. Therefore, the \gls{SD} method has advantages at high $N$ and $\eta_\mathrm{rb}$ and low $M$, where the \gls{TMSE} method is weakest.

\section{Conclusions and Future Work}
\subsection{Conclusion}
In this work, we extended the \gls{SD} to a higher performing low complexity \gls{ISD} by setting the subsurfaces sequentially from the weakest user to the strongest so that the users with the largest impact on total \gls{SNR} benefit from the most prior knowledge. We then proposed the \gls{CISD}, where the \gls{ISD} is repeated until the \gls{SNR} increases by less than a specified threshold. We compared the performance of the \gls{SD}, \gls{ISD} and \gls{CISD}, and found that the iterative methods outperform the \gls{SD}. The largest benefit from the \gls{CISD} occurs at high $N$ where there are more elements that can be further improved with increased knowledge. We also derive the exact closed form expression for the mean \gls{SNR} of the \gls{RIS} \gls{SD} where spatially correlated Ricean fading is assumed for all channels, leading to deeper insights into system performance. This is a very general analysis, and thus we show that the expression can be simplified to verify the results in \cite{inwood_phase_2023} for correlated Rayleigh \gls{UE}-\gls{BS} and \gls{UE}-\gls{RIS} channels and a \gls{LoS} \gls{RIS}-\gls{BS} channel. We showed that even while having significantly less \gls{BW} per user, the \gls{SD} outperformed the much more complex \gls{TMSE} method for the realistic scenarios when a channel is strongly \gls{LoS} or users are clustered together.

\subsection{Future Work}
This work opens up several avenues for future investigation. Firstly, this work focuses on an \gls{UL} system. The proposed methods are also applicable to \gls{DL} operation, and under the assumption of perfect \gls{CSI}, the derived analytical expressions remain structurally unchanged. This assumption is reasonable in the \gls{UL}, as the required effective channel for \gls{RIS} phase optimization can be directly inferred from received pilot signals. However, this assumption is less valid in the \gls{DL}. Relaxing this assumption would require explicitly modeling uncertainty in the channels used for \gls{RIS} phase design, which would modify the analytical evaluation of the mean SNR. Developing such an analysis constitutes an important direction for future work.

Another important direction for future work is the optimization of the number of \gls{RIS} elements allocated to each \gls{UE}. This allocation could be performed either independently of, or jointly with, the \gls{RIS} phase selection methods proposed in this work. The optimization could be based on a range of criteria, such as each \gls{UE}'s channel strength, traffic demand, throughput maximization, or fairness objectives. Optimally partitioning the RIS elements among users has the potential to yield significant performance gains and is therefore a promising topic for further research.

\section*{Appendix A \\ Derivation of Mean SNR}
We compute the mean \gls{SNR} by taking the expectation of each term in (\ref{eq:SNRterms}). The first two terms, $\mathbb{E}[\mathbf{h}_\mathrm{d}^{(k)\dagger}\mathbf{h}_\mathrm{d}^{(k)}]$ and $\mathbb{E}[\mathbf{h}_\mathrm{d}^{(k)}\mathbf{f}_k]$, were derived in \cite{singh_optimal_2022} with their results listed in (\ref{eq:Ehdhd}) and (\ref{eq:Ehdf}). There are four remaining terms to derive.
\vskip 0.1in
\textit{Term 1}: $\mathbb{E}[\mathbf{h}_\mathrm{d}^{(k)\dagger}\mathbf{g}_k]$ \\
As the scattered terms are uncorrelated and hence zero mean, 
\begin{align}
    \mathbb{E}[\mathbf{h}_\mathrm{d}^{(k)\dagger}\mathbf{g}_k] &= \mathbf{h}_\mathrm{d}^{(k,\mathrm{LoS})\dagger}\sum_{s\neq k}\mathbf{H}_{\mathrm{rb},s}^{(k,\mathrm{LoS})}\mathbb{E}[\mathbf{\Phi}_s]\mathbf{h}_{\mathrm{ur},s}^{(k,\mathrm{LoS})}. \notag \\
    \intertext{Using the definitions for LoS terms in Sec. \ref{sec:channel},}
    \mathbb{E}[\mathbf{h}_\mathrm{d}^{(k)\dagger}\mathbf{g}_k] &= \sqrt{\beta^{(k)}_\mathrm{d}\beta_\mathrm{rb}^{(k)}\beta_\mathrm{ur}^{(k)}}\eta^{(k)}_\mathrm{d}\eta_\mathrm{rb}^{(k)}\eta_\mathrm{ur}^{(k)}\mathbf{a}_\mathrm{d}^{(k)\dagger}\mathbf{a}_\mathrm{b}^{(k)} \notag\\  &\qquad  \times \sum_{s\neq k}\mathbf{a}_{\mathrm{r},s}^{(k)\dagger}\mathbf{C}(s)\mathbf{a}_{\mathrm{ur},s}^{(k)},
\end{align}
where $\mathbf{C}(s)$ is as given in \ref{eq:Cx} and derived in Appendix B.
\vskip 0.1in
\textit{Term 2}: $\mathbb{E}[\mathbf{f}_k^\dagger\mathbf{g}_k]$ \\
Expanding out $\mathbf{f}_k$ and $\mathbf{g}_k$ gives
\begin{equation}
    \mathbb{E}[\mathbf{f}_k^\dagger\mathbf{g}_k] = \mathbb{E}\bigg[\mathbf{h}_{\mathrm{ur},k}^{(k)\dagger}\mathbf{\Phi}_k^\dagger\mathbf{H}_{\mathrm{rb},k}^{(k)\dagger}\sum_{s\neq k}\mathbf{H}_{\mathrm{rb},s}^{(k)}\mathbf{\Phi}_s\mathbf{h}_{\mathrm{ur},s}^{(k)}\bigg], \notag \\
\end{equation}
\begin{multline}
    \mathbb{E}[\mathbf{f}_k^\dagger\mathbf{g}_k] = \sum_{s\neq k}\mathbb{E}\left[\nu_k^*\right]\sum_{i=1}^{N_k}\sum_{i=1}^{N_k}\mathbb{E}\left[|\mathbf{h}_{\mathrm{ur},k,i}^{(k)}|\mathbf{h}_{\mathrm{ur},s,j}^{(k)}\right] \\ \times\mathbb{E}\left[(\mathbf{H}_{\mathrm{rb},k}^{(k)\dagger}\mathbf{H}_{\mathrm{rb},s}^{(k)})_{ij}\right](\mathbf{C}(s))_j\mathrm{e}^{-j\angle\mathbf{a}_{\mathrm{r},k,i}^{(k)}}. \label{eq:Efkgkterms}
\end{multline}
From (\ref{eq:Enux}) and (\ref{eq:kabhd}) in Appendix B, 
\begin{equation}
    \mathbb{E}[\nu_k^*] = \frac{\sqrt{\pi\kappa_\mathrm{d}^{(k)}}\mathbf{a}_\mathrm{d}^{(k)\dagger}\mathbf{a}_\mathrm{b}^{(k)}}{2\sqrt{\mathbf{a}_\mathrm{b}^{(k)\dagger}\mathbf{R}_\mathrm{d}^{(k)}\mathbf{a}_\mathrm{b}^{(k)}}}{}_1\!F_1\!\left(\frac{1}{2},2,\frac{-|\mathbf{a}_\mathrm{b}^{(k)\dagger}\mathbf{a}_\mathrm{d}^{(k)}|^2\kappa_\mathrm{d}^{(k)}}{\mathbf{a}_\mathrm{b}^{(k)\dagger}\mathbf{R}_\mathrm{d}^{(k)}\mathbf{a}_\mathrm{b}^{(k)}}\!\right)\!\!.\label{eq:vkconj}
\end{equation}
As $\mathbf{h}_{\mathrm{ur}}^{(k)}$ is a correlated Ricean vector, we can write  $\mathbf{h}_{\mathrm{ur},k,i}^{(k)} = \mathbf{a}_1 + b\mathbf{u}_1$ and $\mathbf{h}_{\mathrm{ur},s,j}^{(k)} = \mathbf{a}_2 + b_2\mathbf{u}_2$, where $\mathbf{a}_1$ and $\mathbf{a}_2$ represent the \gls{LoS} components, $b_1$ and $b_2$ represent the scaling of the scattered terms $\mathbf{u}_1$ and $\mathbf{u}_2$. The Gaussian term of any element can be written as a scaled version of another with noise, so $\mathbf{a}_2=\rho\mathbf{u}_1 + e$ and $\mathbf{h}_{\mathrm{ur},s,j}^{(k)} = \mathbf{a}_2 + b_2(\rho\mathbf{u}_1 + e)$, where $e$ represents noise and $\rho$ represents the spatial correlation between the elements. Returning to the standard channel definitions in  (\ref{eq:hruexpand}) allows us to rewrite $\mathbb{E}\left[|\mathbf{h}_{\mathrm{ur},k,i}^{(k)}|\mathbf{h}_{\mathrm{ur},s,j}^{(k)}\right]$ as
\begin{align}
    \mathbb{E}&\left[|\mathbf{h}_{\mathrm{ur},k,i}^{(k)}|\mathbf{h}_{\mathrm{ur},s,j}^{(k)}\right]\!=\!\sqrt{\frac{\beta_\mathrm{ur}^{(k)}\kappa_\mathrm{ur}^{(k)}}{(1+\kappa_\mathrm{ur}^{(k)})}}\!\!\left(\!\mathbf{a}_{\mathrm{ur},s,j}^{(k)}\!-\!\mathbf{R}_{\mathrm{ur},k,s,i,j}^{(k)}\mathbf{a}_{\mathrm{ur},k,i}^{(k)}\!\right)\!\notag \\ &\times\mathbb{E}\left[|\mathbf{h}_{\mathrm{ur},k,i}^{(k)}|\right] + \mathbf{R}_{\mathrm{ur},k,s,i,j}^{(k)}\mathbb{E}\left[|\mathbf{h}_{\mathrm{ur},k,i}^{(k)}|^2\mathrm{e}^{j\angle\mathbf{h}_{\mathrm{ur},k,i}^{(k)}}\right]\!.
\end{align}
From (4.1) in \cite{miller_complex_1974},
\begin{align}
    \mathbb{E}\left[|\mathbf{h}_{\mathrm{ur},k,i}^{(k)}|\right]\!\!=\!\!\sqrt{\!\frac{\beta_\mathrm{ur}^{(k)}}{(1+\kappa_\mathrm{ur}^{(k)})}}\mathrm{e}^{-\kappa_\mathrm{ur}^{(k)}}\!\Gamma\!\left(\!\frac{3}{2}\!\right)\!\!{}_1F_1\!\!\left(\!\frac{3}{2},1,\kappa_\mathrm{ur}^{(k)}\!\!\right)\!\!.
\end{align}
Using the \gls{PDF} for a complex random variable (RV) in (3.4) of \cite{miller_complex_1974} and letting $r=|\mathbf{h}_{\mathrm{ur},k,i}^{(k)}|$, $\theta=\angle\mathbf{h}_{\mathrm{ur},k,i}^{(k)}$ and $a = \angle\mathbf
    {a}_{\mathrm{ur},k,i}^{(k)}$,
\begin{align}
    &\mathbb{E}\left[r^2\mathrm{e}^{j\theta}\right] = \int_0^\infty\!\!\!\!\int_0^{2\pi}r^2f^*(r,\theta)\,d\theta \,dr, \notag \\ 
    & = \frac{(1+\kappa_\mathrm{ur}^{(k)})\mathrm{e}^{-\kappa_\mathrm{ur}^{(k)}}}{\pi\beta_\mathrm{ur}^{(k)}}\!\int_{0}^\infty \!\!\!r^3\mathrm{exp}\left(-\frac{(1+\kappa_\mathrm{ur}^{(k)})r^2}{\beta_\mathrm{ur}^{(k)}}\right) \notag \\
    &\times \int_0^{2\pi}\!\!\!\!\mathrm{e}^{j\theta}\mathrm{exp}\!\!\left(\!\!2\sqrt{\frac{\kappa_\mathrm{ur}^{(k)}\!\!\left(1+\kappa_\mathrm{ur}^{(k)}\right)}{\beta_\mathrm{ur}^{(k)}}}r\cos(\theta\!-\!a)\!\!\right)d\theta dr\!.
\end{align}
Using Euler's formula and the difference of cosines relationship in (1.313.5) and then the integral result in (3.937.2) of \cite{gradshteyn_table_1980}, the inner integral can be evaluated to give
\begin{align}
    \mathbb{E}\left[r^2\mathrm{e}^{j\theta}\right]&= \frac{2(1+\kappa_\mathrm{ur}^{(k)})\mathrm{e}^{-\kappa_\mathrm{ur}^{(k)}}}{\beta_\mathrm{ur}^{(k)}}\sqrt{\mathrm{exp}(2j\angle a)}\int_0^\infty\!\!\!r^3\notag \\ &\times\!\exp\!\left(\!-\frac{(1+\kappa_\mathrm{ur}^{(k)})r^2}{\beta_\mathrm{ur}^{(k)}}\!\right) \!\!I_1\!\!\left(\sqrt{\frac{4\kappa_\mathrm{ur}^{(k)}(\kappa_\mathrm{ur}^{(k)}+1)}{\beta_\mathrm{ur}^{(k)}}}\right)\!\!dr\!.\notag
\end{align}
Rewriting the modified Bessel function as in (8.406.3) and integrating according to (6.631.1) in \cite{gradshteyn_table_1980} gives
\begin{equation}
    \mathbb{E}\left[r^2\mathrm{e}^{j\theta}\right]\!=\! \frac{\beta_\mathrm{ur}^{(k)}\!\sqrt{\!\kappa_\mathrm{ur}^{(k)}}\mathrm{e}^{j\angle a- \kappa_\mathrm{ur}^{(k)}}\Gamma\!\left(\frac{5}{2}\right)}{(1+\kappa_\mathrm{ur}^{(k)})}{}_1F_1\!\!\left(\!\frac{5}{2},2,\kappa_\mathrm{ur}^{(k)}\!\right)\!.
\end{equation}
Therefore,
\begin{align}
    &\mathbb{E}\left[|\mathbf{h}_{\mathrm{ur},k,i}^{(k)}|\mathbf{h}_{\mathrm{ur},s,j}^{(k)}\right]\!=\!\frac{\beta_\mathrm{ur}^{(k)}\!\sqrt{\!\pi\kappa_\mathrm{ur}^{(k)}}}{2(1+\kappa_\mathrm{ur}^{(k)})}\!\Bigg[\!\Big(\!\mathbf{a}_{\mathrm{ur},s,j}^{(k)}\!-\!\mathbf{R}_{\mathrm{ur},k,s,i,j}^{(k)}\mathbf{a}_{\mathrm{ur},k,i}^{(k)}\!\Big)\notag \\ &\!\!\times\! {}_{\!1}{\!F}_{\!1}\!\!\left(\!\frac{3}{2},1,\kappa_\mathrm{ur}^{(k)}\!\right)\!\!+\!\frac{3}{2}\mathrm{e}^{j\angle\mathbf{a}_{\mathrm{ur},k,i}^{(k)}}\mathbf{R}_{\mathrm{ur},k,s,i,j}^{(k)}\,{}_{\!1}{\!F}_{\!1}\!\!\left(\!\frac{5}{2},\!1,\!\kappa_\mathrm{ur}^{(k)}\!\!\right)\!\!\Bigg]\!.\!\!\!\label{eq:Eabshruhru}
\end{align}
Expanding $\mathbf{H}_\mathrm{rb}^{(k)}$ into the form seen in (\ref{eq:Hrbexpand}), taking the expectation of each term and removing zero mean cross products,
\begin{align}
    \mathbb{E}\left[(\mathbf{H}_{\mathrm{rb},k}^{(k)\dagger}\mathbf{H}_{\mathrm{rb},s}^{(k)})_{ij}\right] \!&= \!\frac{M\beta_\mathrm{rb}^{(k)}}{1+\kappa_\mathrm{rb}^{(k)}}\Big(\kappa_\mathrm{rb}^{(k)}\mathbb{E}\left[\Tilde{\mathbf{H}}_{\mathrm{rb},k,i}^{(k,\mathrm{LoS})\dagger}\Tilde{\mathbf{H}}_{\mathrm{rb},s,j}^{(k,\mathrm{LoS})}\right] \notag \\ &\quad+ \mathbb{E}\left[\Tilde{\mathbf{H}}_{\mathrm{rb},k,i}^{(k,\mathrm{SC})\dagger}\Tilde{\mathbf{H}}_{\mathrm{rb},s,j}^{(k,\mathrm{SC})}\right]\Big), \label{eq:HbrkHbrsij} \notag \\
    &=\!\frac{M\beta_\mathrm{rb}^{(k)}}{1+\kappa_\mathrm{rb}^{(k)}}\Big(\!\kappa_\mathrm{rb}^{(k)}\mathbf{a}_{\mathrm{r},k,i}^{(k)}\mathbf{a}_{\mathrm{r},s,j}^{(k)*}\!+\!\mathbf{R}_{\mathrm{r},k,s,i,j}^{(k)}\!\Big)\!.
\end{align}
Finally, $\mathbf{C}(s)$ is defined in (\ref{eq:Cx}) and will be derived in Appendix B. Substituting (\ref{eq:vkconj}), (\ref{eq:Eabshruhru}), (\ref{eq:HbrkHbrsij}) and (\ref{eq:Cx}) into (\ref{eq:Efkgkterms}) gives (\ref{eq:Efkgk}).
\vskip 0.1in 
\textit{Term 3}: $\mathbb{E}[\mathbf{f}_k^\dagger\mathbf{f}_k]$ \\
Expanding out $\mathbf{f}_k$ with $\mathbf{H}_\mathrm{rb}^{(k)}$ split into \gls{LoS} and scattered components (as seen in (\ref{eq:Hrbexpand})) and taking the mean value leaves only the squared terms, due to the independence of the cross product terms. Hence,
\begin{align}
    \mathbb{E}[&\mathbf{f}_k^\dagger\mathbf{f}_k] = \mathbb{E}\left[\mathbf{h}_{\mathrm{ur},k}^{(k)\dagger}\mathbf{\Phi}_k^\dagger\mathbf{H}_{\mathrm{rb},k}^{(k,\mathrm{LoS})\dagger}\mathbf{H}_{\mathrm{rb},k}^{(k,\mathrm{LoS})}\mathbf{\Phi}_k\mathbf{h}_{\mathrm{ur},k}^{(k)}\right] \notag \\ &+ \mathbb{E}\left[\mathbf{h}_{\mathrm{ur},k}^{(k)\dagger}\mathbf{\Phi}_k^\dagger\mathbf{H}_{\mathrm{rb},k}^{(k,\mathrm{SC})\dagger}\mathbf{H}_{\mathrm{rb},k}^{(k,\mathrm{SC})}\mathbf{\Phi}_k\mathbf{h}_{\mathrm{ur},k}^{(k)}\right]\!=\!T_1\!+\!T_2.
\end{align}
Splitting $T_1$ into a sum of its terms, applying the same process as in (\ref{eq:HbrkHbrsij}) for $\mathbb{E}[(\mathbf{H}_{\mathrm{rb},k}^{(k,\mathrm{LoS})\dagger}\mathbf{H}_{\mathrm{rb},k}^{(k,\mathrm{LoS}})_{ij}]$, replacing $\mathbf{\Phi}_k$ with its expanded form in (\ref{eq:Phi}) and simplifying gives
\begin{align}
    T_1\!=\!M\beta_\mathrm{rb}^{(k)}\eta_\mathrm{rb}^{(k)2}\!\!\left(\!N_k\!+\!\sum_{i=1}^{N_k}\sum_{j\neq i}\mathbb{E}\left[|\mathbf{h}_{\mathrm{ur},k,i}^{(k)}||\mathbf{h}_{\mathrm{ur},k,j}^{(k)}|\right]\!\right)\!. \label{eq:T1}
\end{align}
From \cite{singh_optimal_2022},
\begin{align}
    &\mathbb{E}\!\left[|\mathbf{h}_{\mathrm{ur},k,i}^{(k)}||\mathbf{h}_{\mathrm{ur},k,j}^{(k)}|\right] = \beta_\mathrm{ur}^{(k)}F_R = \frac{\beta_\mathrm{ur}^{(k)}(1-|\rho^{(k)}_{kk}|^2)^2}{(1+\kappa_\mathrm{ur}^{(k)})}\notag \\
    &\!\times\!\exp{\!\left(\!\frac{-2\kappa_\mathrm{ru}^{(k)}(1\!-\!\mu_{cf})}{1\!-\!|\rho^{(k)}_{kk}|^2}\!\right)} \!\!\sum_{m=0}^\infty\sum_{n=0}^m\cos(n\phi_f)\frac{\epsilon_n|\rho^{(k)}_{kk}|^{2m-n}}{m!(m\!-\!n)!(n!)^2}\notag \\
    &\!\times\!\left(\!\frac{\kappa_\mathrm{ru}^{(k)}(1+|\rho^{(k)}_{kk}|^2\!-\!2\mu_{cf}))}{1-|\rho^{(k)}_{kk}|^2}\!\!\right)^{\!\!n}\!\!\Gamma^2\!\left(\!m\!+\!\frac{3}{2}\!\right)\!{}_1F_1^2\Bigg(\!m\!+\!\frac{3}{2}, n\!+\!1, \notag \\ &\quad\frac{\kappa_\mathrm{ru}^{(k)}(1+|\rho^{(k)}_{kk}|^2-2\mu_{cf}))}{1-|\rho^{(k)}_{kk}|^2}\!\Bigg)\!, \label{eq:Ehurihurj}
\end{align}
where $\rho_{kk}^{(k)}=\mathbf{R}^{(k)}_{\mathrm{ur},k,k,i,j}$, $\mu_{cf}\!=\!\rho^{(k)}_{kk}\!\cos\!\left(\!\angle\mathbf{a}_{\mathrm{ur},k,i}^{(k)}\!-\!\angle\mathbf{a}_{\mathrm{ur},k,j}^{(k)}\!\right)$, $\mu_{sf}\!=\! \rho^{(k)}_{kk}\!\sin\!\left(\angle\mathbf{a}_{\mathrm{ur},k,i}^{(k)}\!-\!\angle\mathbf{a}_{\mathrm{ur},k,j}^{(k)}\!\right)$, $\phi_f\!=\!\angle(\!(1\!+\!|\rho_{kk}^{(k)}\!|^2)\mu_c\kappa_\mathrm{ur}^{(k)}\!\!-\!\!2\kappa_\mathrm{ur}^{(k)}\!|\rho_{kk}^{(k)}\!|^2\!\!+\!j(1\!-\!|\rho_{kk}^{(k)}\!|^2)\mu_s\kappa_\mathrm{ur}^{(k)})$, $\epsilon_0 = 1$, and $\epsilon_n = 2$ when $n\geq 1$ as defined in \cite{mendes_general_2007}.

$T_2$ can be found using the same process as for $T_1$, giving
\begin{align}
    T_2 \!&=\! M\beta_\mathrm{rb}^{(k)}\zeta_\mathrm{rb}^{(k)2}\!\left(\!N_k\!+\!\sum_{i=1}^{N_k}\sum_{j\neq i}\mathbb{E}\!\left[|\mathbf{h}_{\mathrm{ur},k,i}^{(k)}||\mathbf{h}_{\mathrm{ur},k,j}^{(k)}|\right]\!\mathbf{A}_{i,j}\!\right)\!,\label{eq:T2}
\end{align}
where
\begin{align}
    \mathbf{A}_{i,j} &= \left(\mathrm{diag}\big(\mathbf{a}_\mathrm{r}^{(k)}\big)^\dagger\mathbf{R}_{\mathrm{r},k,k}^{(k)}\mathrm{diag}\big(\mathbf{a}_\mathrm{r}^{(k)}\big)\right)_{i,j},
\end{align}
and the result from (\ref{eq:Ehurihurj}) is used for $\mathbb{E}[|\mathbf{h}_{\mathrm{ur},k,i}^{(k)}||\mathbf{h}_{\mathrm{ur},k,j}^{(k)}|]$. Combining (\ref{eq:T1}) and (\ref{eq:T2}) gives (\ref{eq:Efkfk}).

\vskip 0.1in
\textit{Term 4}: $\mathbb{E}[\mathbf{g}_k^\dagger\mathbf{g}_k]$\\
Splitting $\mathbb{E}[\mathbf{g}_k^\dagger\mathbf{g}_k]$ into two sums,
\begin{multline}
    \mathbb{E}[\mathbf{g}_k^\dagger\mathbf{g}_k] = \mathbb{E}\Big[\!\!\sum_{s=t\neq k}\!\!\mathbf{h}_{\mathrm{ur},s}^{(k)\dagger}\mathbf{\Phi}_s^\dagger\mathbf{H}_{\mathrm{rb},s}^{(k)\dagger}\mathbf{H}_{\mathrm{rb},s}^{(k)}\mathbf{\Phi}_s\mathbf{h}_{\mathrm{ur},s}^{(k)}\Big] \\ + \mathbb{E}\Big[\!\!\sum_{s\neq t\neq k}\!\!\mathbf{h}_{\mathrm{ur},s}^{(k)\dagger}\mathbf{\Phi}_s^\dagger\mathbf{H}_{\mathrm{rb},s}^{(k)\dagger}\mathbf{H}_{\mathrm{rb},t}^{(k)}\mathbf{\Phi}_t\mathbf{h}_{\mathrm{ur},t}^{(k)}\Big]\!=\!G_1\!+\!G_2. \label{eq:G1G2}
\end{multline}
Rewriting $G_1$ as a sum of its terms and taking the expectation of each independent pair of terms gives
\begin{multline}
     G_1=\!\!\!\sum_{s=t\neq k}\sum_{i=1}^{N_k}\sum_{j=1}^{N_k}\mathbb{E}\Big[\big(\mathbf{H}_{\mathrm{rb},s}^{(k)\dagger}\mathbf{H}_{\mathrm{rb},s}^{(k)}\big)_{i,j}\Big] \\ \times\mathbb{E}\big[\mathbf{\Phi}_{\!s,i}^*\mathbf{\Phi}_{\!s,j}\big]\mathbb{E}\big[\mathbf{h}_{\mathrm{ur},s,i}^{(k)*}\mathbf{h}_{\mathrm{ur},s,j}^{(k)}\big]. \label{eq:G1}
\end{multline}
Following the procedure used for (\ref{eq:HbrkHbrsij}),
\begin{equation}
    \mathbb{E}\left[\big(\mathbf{H}_{\mathrm{rb},s}^{(k)\dagger}\mathbf{H}_{\mathrm{rb},s}^{(k)}\big)_{i,j}\right]\!= \!\frac{M\!\sqrt{\!\beta_\mathrm{rb}^{(k)}}}{1+\kappa_\mathrm{rb}^{(k)}}\!\Big(\!\kappa_\mathrm{rb}^{(k)}\mathbf{a}_{\mathrm{r},s,i}^{(k)}\mathbf{a}_{\mathrm{r},s,j}^{(k)*}\!+\!\mathbf{R}_{\mathrm{r},s,s,i,j}^{(k)}\!\Big)\!. \label{eq:HbrsiHbrsj}
\end{equation}
Expanding out $\mathbf{\Phi}_s$ and simplifying the expression gives
\begin{equation}
    \mathbb{E}[\mathbf{\Phi}_{\!s,i}^*\mathbf{\Phi}_{\!s,j}]= \mathrm{e}^{-j\angle\mathbf{a}_{\mathrm{r},s,i}^{(s)}}\mathrm{e}^{j\angle\mathbf{a}_{\mathrm{r},s,j}^{(s)}}\mathbb{E}\!\left[\mathrm{e}^{j\angle\mathbf{h}_{\mathrm{ur},s,i}^{(s)}}\mathrm{e}^{-j\angle\mathbf{h}_{\mathrm{ur},s,j}^{(s)}}\right]\!, \label{eq:EPhiPhi}
\end{equation}
where $G_R\!=\!\mathbb{E}[\mathrm{e}^{j\angle\mathbf{h}_{\mathrm{ur},s,i}^{(s)}}\mathrm{e}^{-j\angle\mathbf{h}_{\mathrm{ur},s,j}^{(s)}}]$. Appendix C will derive $G_R$, with the final result stated in (\ref{eq:GR_body}). Expanding $\mathbf{h}_\mathrm{ur}^{(k)}$ into the form seen in (\ref{eq:hruexpand}), taking the expectation of each term and removing the zero mean cross product terms gives
\begin{equation}
    \mathbb{E}\big[\mathbf{h}_{\mathrm{ur},s,i}^{(k)*}\mathbf{h}_{\mathrm{ur},s,j}^{(k)}\big]\!= \!\frac{\!\sqrt{\!\beta_\mathrm{ur}^{(k)}}}{1+\kappa_\mathrm{ur}^{(k)}}\!\Big(\!\kappa_\mathrm{ur}^{(k)}\mathbf{a}_{\mathrm{ur},s,i}^{(k)*}\mathbf{a}_{\mathrm{ur},s,j}^{(k)}\!+\!\mathbf{R}_{\mathrm{ur},s,s,i,j}^{(k)}\!\Big)\!. \label{eq:hrusihrusj}
\end{equation}
Combining (\ref{eq:HbrsiHbrsj}), (\ref{eq:EPhiPhi}) and (\ref{eq:hrusihrusj}) gives (\ref{eq:G1}). As for $G_1$, rewriting $G_2$ as a sum of its terms and taking the expectation of each independent group gives
\begin{multline}
    G_2=\sum_{s \neq t\neq k}\!\sum_{i=1}^{N_k}\sum_{j=1}^{N_k}\mathbb{E}\left[\big(\mathbf{H}_{\mathrm{rb},s}^{(k)\dagger}\mathbf{H}_{\mathrm{rb},t}^{(k)}\big)_{i,j}\right] \\ \times\mathbb{E}\big[\mathbf{\Phi}_{s,i}^*\big]\mathbb{E}\big[\mathbf{\Phi}_{t,j}\big]\mathbb{E}\big[\mathbf{h}_{\mathrm{ur},s,i}^{(k)*}\mathbf{h}_{\mathrm{ur},t,j}^{(k)}\big]. \label{eq:G2}
\end{multline}
Again, from (\ref{eq:HbrkHbrsij}),
\begin{equation}
    \left[\big(\mathbf{H}_{\mathrm{rb},s}^{(k)\dagger}\mathbf{H}_{\mathrm{rb},t}^{(k)}\big)_{i,j}\right]\!= \!\frac{M\!\sqrt{\!\beta_\mathrm{rb}^{(k)}}}{1+\kappa_\mathrm{rb}^{(k)}}\!\Big(\!\kappa_\mathrm{rb}^{(k)}\mathbf{a}_{\mathrm{r},s,i}^{(k)}\mathbf{a}_{\mathrm{r},t,j}^{(k)*}\!+\!\mathbf{R}_{\mathrm{r},s,t,i,j}^{(k)}\!\Big)\!. \label{eq:HbrsiHbrtj}
\end{equation}
As will be shown in Appendix B,
\begin{multline}
    \!\!\!\!\!(\mathbf{C}^*\!(s)\!)_{i,i}\!=\!\mathbb{E}\big[\mathbf{\Phi}^*_{s,i}\big]\!=\! \frac{\pi\sqrt{\!\kappa_\mathrm{d}^{(s)}\!\kappa_\mathrm{ur}^{(s)}}\mathbf{a}_\mathrm{d}^{(s)\dagger}\mathbf{a}_\mathrm{b}^{(s)}\!}{4\sqrt{\mathbf{a}_\mathrm{b}^{(s)\dagger}\mathbf{R}_{\mathrm{d}}^{(s)}\mathbf{a}_\mathrm{b}^{(s)}}}\mathrm{e}^{j\left(\!\angle\mathbf{a}_{\mathrm{ur},s,i}^{(s)}\!-\!\angle\mathbf{a}_{\mathrm{r},s,i}^{(s)}\right)} \\ \times{}_1F_1\!\!\left(\!\frac{1}{2},\!2,\!-\kappa_\mathrm{ur}^{(s)}\!\!\right)\!\!{}_1F_1\!\!\left(\!\!\frac{1}{2},\!2,\!\frac{-|\mathbf{a}_\mathrm{b}^{(s)\dagger}\mathbf{a}_\mathrm{d}^{(s)}\!|^2\kappa_\mathrm{d}^{(s)}}{\mathbf{a}_\mathrm{b}^{(s)\dagger}\mathbf{R}_{\mathrm{d}}^{(s)}\mathbf{a}_\mathrm{b}^{(s)}}\!\right)\!\!,\! \label{eq:Csi} 
\end{multline}
\begin{multline}
    \!\!\!\!\!\!(\mathbf{C}(t)\!)_{j,j}\!=\!\mathbb{E}\big[\mathbf{\Phi}_{t,j}\big]\!=\! \frac{\pi\sqrt{\kappa_\mathrm{d}^{(t)}\kappa_\mathrm{ur}^{(t)}}\mathbf{a}_\mathrm{b}^{(t)\dagger}\mathbf{a}_\mathrm{d}^{(t)}}{4\sqrt{\mathbf{a}_\mathrm{b}^{(t)\dagger}\mathbf{R}_{\mathrm{d}}^{(t)}\mathbf{a}_\mathrm{b}^{(t)}}}\mathrm{e}^{j\left(\angle\mathbf{a}_{\mathrm{r},t,j}^{(t)}\!-\!\angle\mathbf{a}_{\mathrm{ur},t,j}^{(t)}\right)}\!  \\ \times{}_1\!F_1\!\!\left(\!\frac{1}{2},\!2,\!-\kappa_\mathrm{ur}^{(t)}\!\!\right)\!\!{}_1F_1\!\!\left(\!\!\frac{1}{2},\!2,\!\frac{-|\mathbf{a}_\mathrm{b}^{(t)\dagger}\mathbf{a}_\mathrm{d}^{(t)}|^2\kappa_\mathrm{d}^{(t)}}{\mathbf{a}_\mathrm{b}^{(t)\dagger}\mathbf{R}_{\mathrm{d}}^{(t)}\mathbf{a}_\mathrm{b}^{(t)}}\!\right)\!\!. \label{eq:Ctj}
\end{multline}
Finally, using the same process as for (\ref{eq:hrusihrusj}),
\begin{equation}
    \mathbb{E}\big[\mathbf{h}_{\mathrm{ur},s,i}^{(k)*}\mathbf{h}_{\mathrm{ur},t,j}^{(k)}\big]\!= \!\frac{\!\sqrt{\!\beta_\mathrm{ur}^{(k)}}}{1+\kappa_\mathrm{ur}^{(k)}}\!\Big(\!\kappa_\mathrm{ur}^{(k)}\mathbf{a}_{\mathrm{ur},s,i}^{(k)*}\mathbf{a}_{\mathrm{ur},t,j}^{(k)}\!+\!\mathbf{R}_{\mathrm{ur},s,t,i,j}^{(k)}\!\Big)\!. \label{eq:hrusihrutj}
\end{equation}
Therefore combining (\ref{eq:HbrsiHbrtj}), (\ref{eq:Csi}), (\ref{eq:Ctj}) and (\ref{eq:hrusihrutj}) gives (\ref{eq:G2}), and combining (\ref{eq:G1}) and (\ref{eq:G2}) gives (\ref{eq:G1G2}).

\section*{Appendix B \\ Derivation of $\mathbf{C}(x) = \mathbb{E}[\mathbf{\Phi}_x]$}
Let $\mathbf{C}(x) = \mathbb{E}[\mathbf{\Phi}_x]$. From the definition of $\mathbf{\Phi}_x$ in (\ref{eq:Phi}),
\begin{equation}
    \mathbb{E}[\mathbf{\Phi}_x] = \mathbb{E}[\nu_x]\mathrm{diag}\big(\mathrm{e}^{j\angle\mathbf{a}^{(x)}_{\mathrm{r},x}}\big)\mathbb{E}\Big[\mathrm{diag}\big(\mathrm{e}^{-j\angle\mathbf{h}_{\mathrm{ur},x}^{(x)}}\big)\Big]. \label{eq:EPhix}
\end{equation}
From (4.12) in \cite{miller_complex_1974},
\begin{equation}
    \mathbb{E}[\mathrm{e}^{-j\angle\mathbf{h}_{\mathrm{ur},x}^{(x)}}]\!\!=\!\!\frac{\sqrt{\pi\kappa_\mathrm{ur}^{(x)}}\!}{2}\mathrm{diag}\!\left(\!\mathrm{e}^{-j\angle\mathbf{a}_{\mathrm{ur},x}^{(x)}}\!\right)\!\!{}_1{\!F}_{\!1}\!\!\left(\!\frac{3}{2},\!2,\!-\kappa_\mathrm{ur}^{(x)}\!\!\right)\!\!,\! \label{eq:Eejhrux}
\end{equation}
\begin{equation}
    \mathbb{E}[\nu_x]\!\!=\!\!\frac{\sqrt{\!\pi\kappa_\mathrm{abhd}^{(x)}}}{2}\mathrm{diag}\!\left(\!\mathrm{e}^{j\angle\mathbf{a}_{\mathrm{b}}^{(x)\dagger}\mathbf{a}_\mathrm{d}^{(x)}}\!\right)\!{}_1\!F_1\!\!\left(\!\frac{3}{2},2,-\kappa_\mathrm{abhd}^{(x)}\!\!\right)\!, \label{eq:Enux}
\end{equation}
where $\kappa_\mathrm{abhd}^{(x)}$ is the K-factor of  $\mathbf{a}_\mathrm{b}^{(x)\dagger}\mathbf{h}_\mathrm{d}^{(x)}$ and 
\begin{equation}
    \kappa_\mathrm{abhd}^{(x)} = \kappa_\mathrm{d}^{(x)}\frac{|\mathbf{a}_\mathrm{b}^{(x)\dagger}\mathbf{a}_\mathrm{d}^{(x)}|^2}{\mathbf{a}_\mathrm{b}^{(x)\dagger}\mathbf{R}_\mathrm{d}^{(x)}\mathbf{a}_\mathrm{b}^{(x)}}. \label{eq:kabhd}
\end{equation}
Therefore, combining (\ref{eq:EPhix}), (\ref{eq:Eejhrux}) and (\ref{eq:Enux}) gives (\ref{eq:Cx}).

\section*{Appendix C \\ Derivation of $G_R = \mathbb{E}[\mathrm{e}^{-j\phi_i}\mathrm{e}^{j\phi_j}]$}
Let $G_R = \mathbb{E}\left[\mathrm{e}^{j\angle\mathbf{h}_{\mathrm{ur},s,i}^{(s)}}\mathrm{e}^{-j\angle\mathbf{h}_{\mathrm{ur},s,j}^{(s)}}\right] = \mathbb{E}[\mathrm{e}^{-j\phi_i}\mathrm{e}^{j\phi_j}]$. Using the joint \gls{PDF} for two complex \glspl{RV} given in (24a) of \cite{mendes_general_2007},
\begin{equation}
    G_R \!=\!\!\!\int_0^\infty\!\!\!\!\int_0^\infty\!\!\!\!\int_{-\pi}^{\pi}\!\int_{-\pi}^{\pi}\!\!\!\!\mathrm{e}^{\!-j\phi_i}\mathrm{e}^{j\phi_j}\!f^*\!\!\left(r_{\!i},r_{\!j},\mathrm{e}^{j\phi_i}\!,\mathrm{e}^{j\phi_j}\!\right)\!d{\phi_i}d{\phi_j}d{r_{\!i}}d{r_{\!j}}. \notag
\end{equation}
Applying the variable transformations $\phi_i=\phi_i-\angle(1-\mu_c-j\mu_s)-\angle{\mathbf{a}^{(s)}_{\mathrm{ur},s,i}}$ and $\phi_j= \phi_j-\angle(1-\mu_c+j\mu_s)-\angle{\mathbf{a}^{(s)}_{\mathrm{ur},s,j}}$,
\begin{align}
    &G_R=\!\frac{\left(\!1\!+\!\kappa_\mathrm{ur}^{(s)}\!\right)^{\!\!2}\!\mathrm{e}^{j\!\left(\!\angle(\!1-\mu_c+j\mu_s\!)+\!\angle\mathbf{a}^{(s)}_{\mathrm{ur},s,j}\!\right)}}{\!\pi^2\!\!\left(\!1\!-\!\rho_{ss}^{(s)2}\!\right)\!\mathrm{e}^{j\!\left(\!\angle(\!1-\mu_c-j\mu_s\!)+\!\angle\mathbf{a}^{(s)}_{\mathrm{ur},s,i}\!\right)}\!\!}\!\exp\!\!\left(\!\!\frac{-2\kappa_\mathrm{ur}^{(s)}\!(\!1\!-\!\mu_c)}{1\!-\!\rho_{ss}^{(s)2}}\!\!\right)\! \notag \\ &\times\!\!\!\int_0^\infty\!\!\!\!\int_0^\infty\!\!\!r_{\!i}r_{\!j}\!\exp\!\!\left(\!\!\frac{-\big(1\!+\!\kappa_\mathrm{ur}^{(s)2}\big)}{1-\rho_{ss}^{(s)}}\!\!\left(r_i^2\!+\!r_j^2\right)\!\!\right)\!\!\int_{-\pi}^{\pi}\!\int_{-\pi}^{\pi}\!\!\mathrm{e}^{-j\phi_i}\mathrm{e}^{j\phi_{\!j}} \notag \\ &\times\!\mathrm{e}^{\kappa_0r_ir_{\!j}\!\cos(\!\phi_{i}\!-\phi_j\!-\phi\!)}\mathrm{e}^{\zeta\!\sqrt{\kappa}(r_i\cos(\!\phi_i\!)+ r_j\cos(\!\phi_j\!)\!)}d\phi_i d\phi_jdr_idr_j,
\end{align}
where $\rho_{ss}^{(s)} = \mathbf{R}^{(s)}_{\mathrm{ur},s,s,i,j}$, $\mu_c = \rho^{(s)}_{ss}\!\cos\!\left(\!\angle\mathbf{a}_{\mathrm{ur},s,j}^{(s)}-\angle\mathbf{a}_{\mathrm{ur},s,i}^{(s)}\!\right)$, $\mu_s = \rho^{(s)}_{ss}\!\sin\!\left(\!\angle\mathbf{a}_{\mathrm{ur},s,j}^{(s)}-\angle\mathbf{a}_{\mathrm{ur},s,i}^{(s)}\!\right)$, $\zeta=\frac{2\sqrt{1+\kappa_\mathrm{ur}^{(s)}}}{1-|\rho^{(s)}_{ss}|^2}$, $\kappa = \kappa_\mathrm{ur}^{(s)}(1+|\rho_{ss}^{(s)}|^2-2\mu_c)$, $\kappa_0 = \frac{2|\rho^{(s)}_{ss}|}{1\!-\!|\rho^{(s)}_{ss}|^2}\!\!\left(\!1\!+\!\kappa_\mathrm{ur}^{(s)}\!\right)$ and $\phi = \angle(1\!-\!\mu_c\!+\!j\mu_s)+\angle\mathbf{a}_{\mathrm{ur},s,j}^{(s)}-\angle(1\!-\!\mu_c\!-\!j\mu_s)-\angle\mathbf{a}_{\mathrm{ur},s,i}^{(s)}+\angle\rho^{(s)}_{ss}$.

We can now integrate with respect to both $\phi_i$ and $\phi_j$. Rewriting $\mathrm{e}^{-j\phi_i}$ and $\mathrm{e}^{j\phi_j}$ using Euler's formula allows the double integral over the phases to be split into four smaller double integrals. Rewriting $\mathrm{e}^{\kappa_0r_ir_j\cos{(\phi_i-\phi_j-\phi)}}$ in the form seen in \cite[8.511.4]{gradshteyn_table_1980} and the resulting $\cos(n\phi_j-n\phi_i-n\phi)$ in the form seen in \cite[1.313.5]{gradshteyn_table_1980}, removing periodic terms with zero-valued integrals, applying the product to sum formula found from \cite[1.313.5]{gradshteyn_table_1980} and finally integrating each of the four double integrals using \cite[3.915.2]{gradshteyn_table_1980}, we get
\begin{align}
    &G_R = \!\frac{2\!\left(\!1\!+\!\kappa_\mathrm{ur}^{(s)}\!\right)^{\!\!2}\!\mathrm{e}^{j\!\left(\!\angle(\!1-\mu_c+j\mu_s\!)+\!\angle\mathbf{a}^{(s)}_{\mathrm{ur},s,j}\!\right)}\!}{\!\left(\!1\!-\!\rho_{ss}^{(s)2}\!\right)\!\mathrm{e}^{j\!\left(\!\angle(\!1-\mu_c-j\mu_s\!)+\!\angle\mathbf{a}^{(s)}_{\mathrm{ur},s,i}\!\right)}}\!\exp\!\!\left(\!\!\frac{-2\kappa_\mathrm{ur}^{(s)}\!(\!1\!-\!\mu_c)}{1\!-\!\rho_{ss}^{(s)2}}\!\!\right)\! \notag \\ &\times \!\!\!\int_0^\infty\!\!\!\!\int_0^\infty\!\!\!r_{\!i}r_{\!j}\!\exp\!\!\left(\!\!\frac{-\big(1\!+\!\kappa_\mathrm{ur}^{(s)2}\big)}{1-\rho_{ss}^{(s)2}}\!\!\left(r_i^2\!+\!r_j^2\right)\!\!\right)\!\!\sum_{n=0}^\infty\epsilon_nI_n(\kappa_0 r_ir_j) \notag \\ &\times \big(J_{n+1}(-j\zeta\sqrt{\kappa} r_i)J_{n+1}(-j\zeta\sqrt{\kappa} r_j)\mathrm{e}^{jn\phi} \notag \\ &+ J_{n-1}(-j\zeta\sqrt{\kappa} r_i)J_{n-1}(-j\zeta\sqrt{\kappa} r_j)\mathrm{e}^{-jn\phi}\big) dr_i dr_j,\!\!
\end{align}
where $\epsilon_0 = 1$ and $\epsilon_n = 2$ when $n \geq 1$. To integrate with respect to $r_i$ and $r_j$, we can rewrite the modified Bessel function $I_n(\kappa_0 r_i r_j)$ as a Bessel function using \cite[8.406.3]{gradshteyn_table_1980}, expand that function to its series form using \cite[8.402]{gradshteyn_table_1980} and integrate each integral using \cite[6.631.1]{gradshteyn_table_1980}, resulting in (\ref{eq:GR_body}).

\bibliographystyle{IEEEtran}
\bibliography{IEEEabrv, journal.bib}
\vspace{-4em}
\begin{IEEEbiography}[{\includegraphics[width=1in,height=1.25in,clip,keepaspectratio]{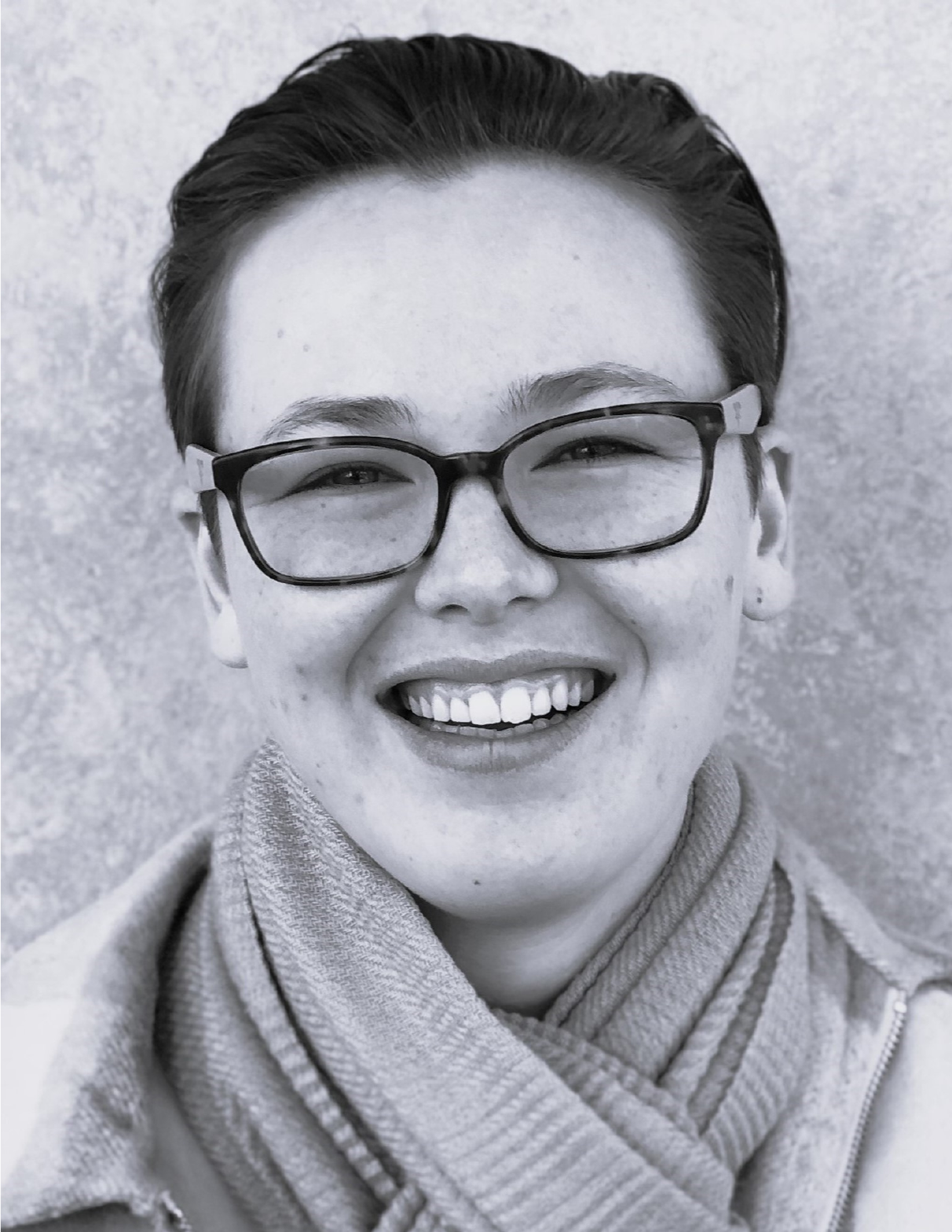}}]%
{Amy S. Inwood} (S'18-M'25) received the B.E (Hons.) and Ph.D. degrees in electrical and electronic engineering from Te Whare Wānanga o Waitaha $|$ University of Canterbury (UC), NZ, in 2021 and 2024, respectively. She is now a Research Fellow at the Centre for Wireless Innovation, Queen's University Belfast, Belfast, U.K. Her research interests include statistical analysis, 5G-6G wireless communications, MIMO, reconfigurable intelligent surfaces and fluid antenna systems.
\end{IEEEbiography}
\vspace{-5em}
\begin{IEEEbiography}[{\includegraphics[width=1in,height=1.25in,clip,keepaspectratio]{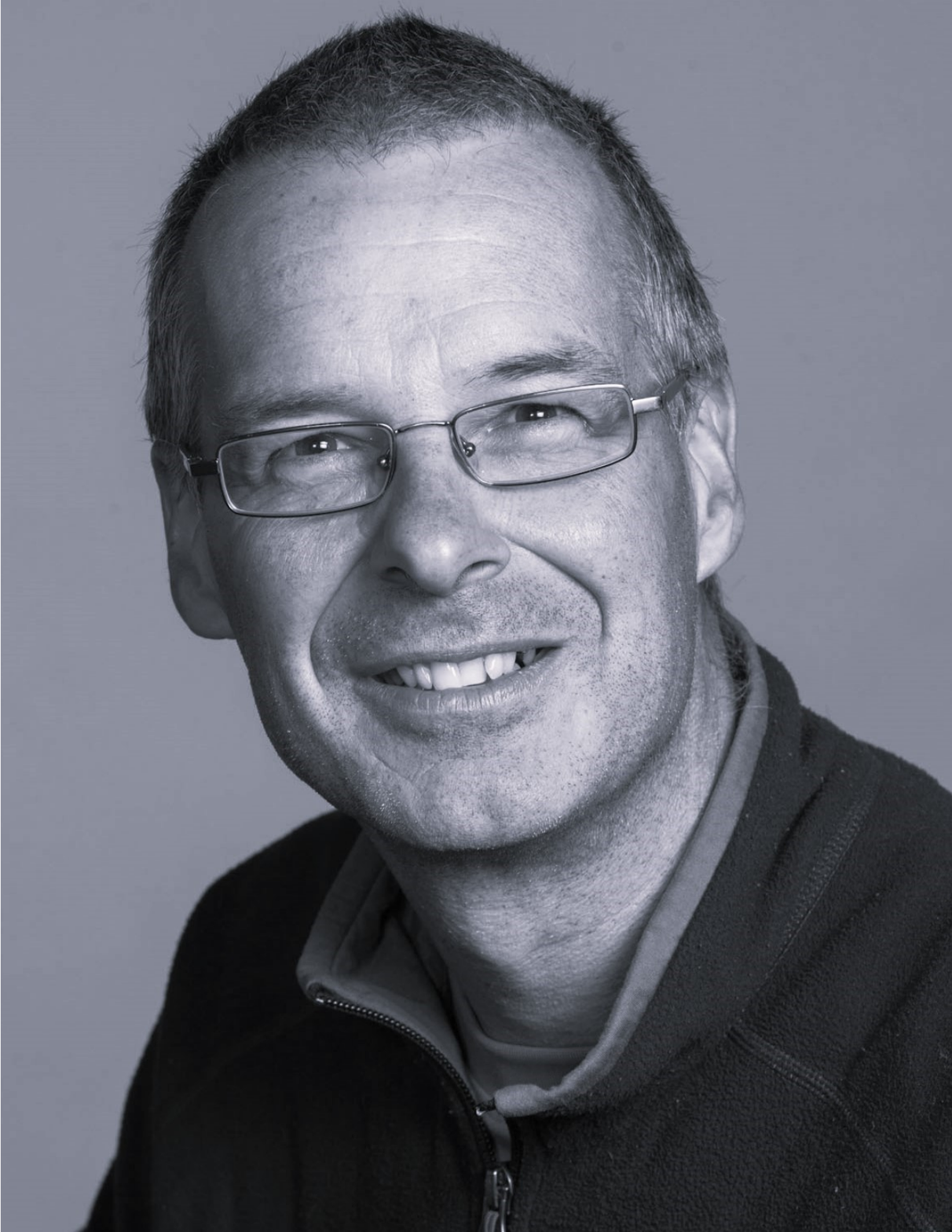}}]%
{Peter J. Smith} (M'93-SM'01-F'15) received the B.Sc degree in Mathematics and the Ph.D degree in Statistics from the University of London, London, U.K., in 1983 and 1988, respectively. In 2015 he joined Victoria University of Wellington as Professor of Statistics. He is an Adjunct Professor in Electrical and Computer Engineering at the University of Canterbury, New Zealand and an Honorary Professor in the School of Electronics, Electrical Engineering and Computer Science, Queens University Belfast. His research interests include the statistical aspects of design, antenna arrays, MIMO, cognitive radio, massive MIMO, mmWave systems, reconfigurable intelligent surfaces and the fusion of radar sensing and communications.
\end{IEEEbiography}
\vspace{-5em}
\begin{IEEEbiography}[{\includegraphics[width=1in,height=1.25in,clip,keepaspectratio]{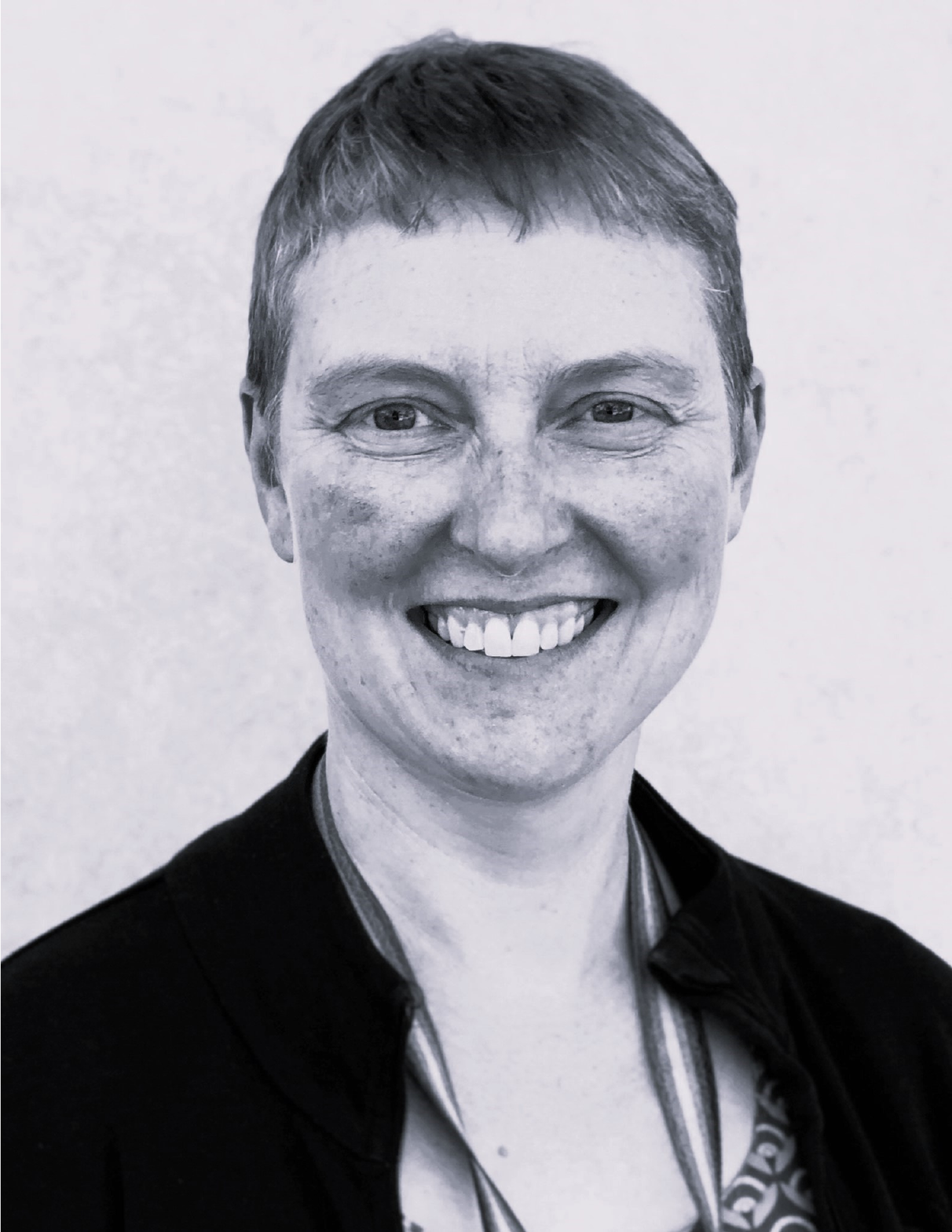}}]%
{Philippa A. Martin} (S'95-M'01-SM'06) received the B.E. (Hons.) and Ph.D. degrees in electrical and electronic engineering from Te Whare Wānanga o Waitaha $|$ University of Canterbury, NZ, in 1997 and 2001, respectively. She is now a Professor there. She is a Fellow of Engineering New Zealand. Her research interests include error correction coding, detection, multi-antenna systems, channel modelling, and 5G-6G wireless communications. She served as an Editor of the IEEE Transactions on Wireless Communications 2005-08, 2014-16 and member of the IEEE Communication Society Board of Governors 2019-21. She is currently on their financial standing committee.
\end{IEEEbiography}
\vspace{-5em}
\begin{IEEEbiography}[{\includegraphics[width=1in,height=1.25in,clip,keepaspectratio]{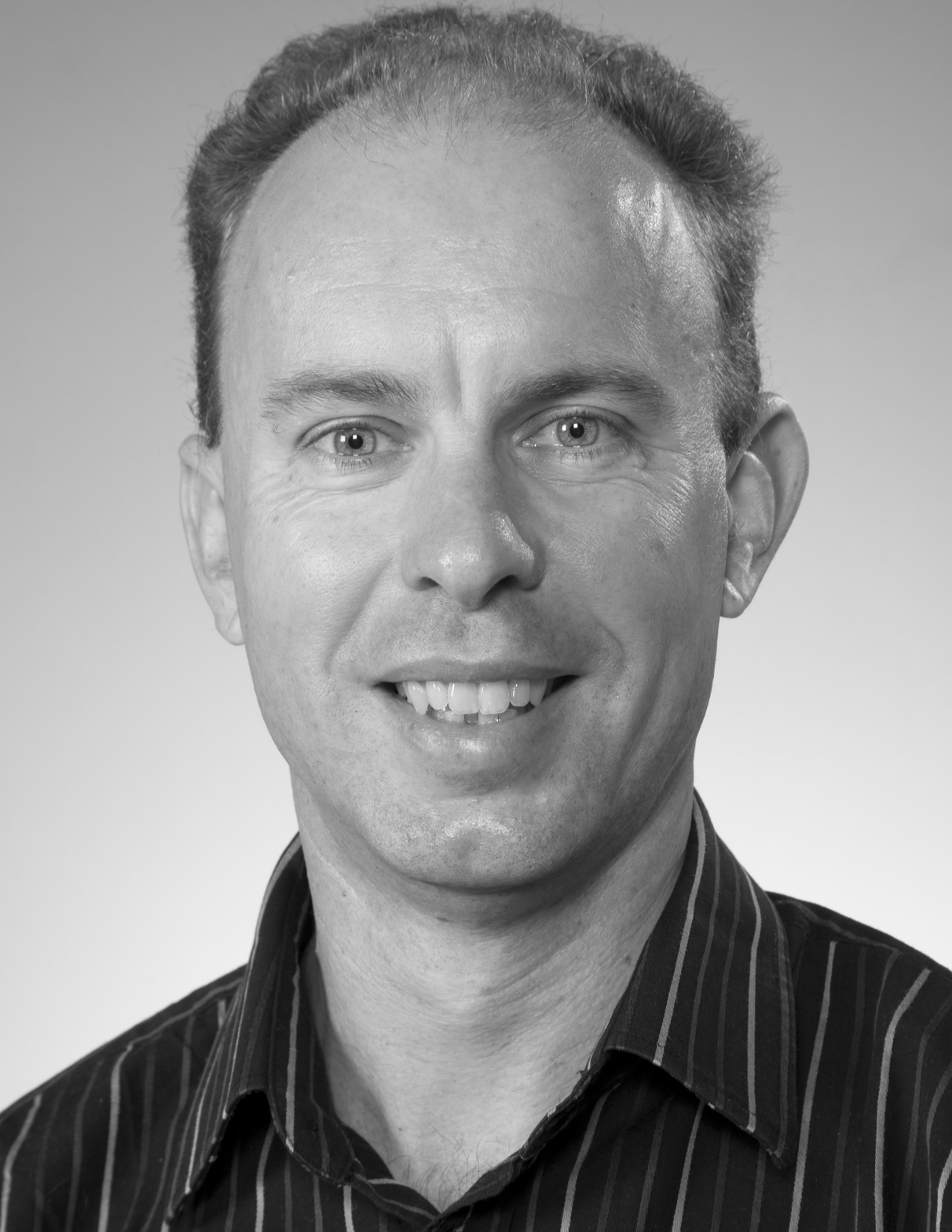}}]{Graeme K. Woodward} (S'94-M'99-SM'05) received the B.Sc., B.E., and Ph.D. degrees from The University of Sydney, NSW, Australia, in 1990, 1992, and 1999, respectively. He has been the Research Leader with the Wireless Research Centre, Te Whare Wānanga o Waitaha $|$ University of Canterbury, NZ, since 2011, and previously the Research Manager of the Telecommunications Research Laboratory, Toshiba Research Europe, contributing to numerous large U.K. and EU projects. His extensive industrial research experience includes pioneering VLSI designs for multi-antenna 3G Packet Access (HSDPA) with Bell Labs Sydney. While holding positions at Agere Systems and LSI Logic, his focus moved to terminal-side algorithms for 3G and 4G (LTE), with an emphasis on low power design. \end{IEEEbiography}

\end{document}